\documentclass[twocolumn,aps,prl,nofootinbib,superscriptaddress,amsmath,amssymb,floatfix]{revtex4-2}
\usepackage{graphicx}
\usepackage{dcolumn}	
\usepackage{bm}			
\usepackage{amsfonts}
\usepackage{xspace}
\usepackage{color}
\usepackage{epstopdf}
\usepackage{multirow}
\usepackage{amsmath}
\usepackage{float}
\usepackage{hyperref}
\usepackage{cleveref}
\usepackage{amsmath}
\usepackage{amssymb}
\usepackage{mathtools}
\usepackage{enumitem}
\usepackage[section]{placeins}

\usepackage{url}
\RequirePackage{hyperref}
\hypersetup{
    breaklinks,
    unicode,
    linktoc=all,
    bookmarksnumbered=true,
    bookmarksopen=true,
    colorlinks,
    linkcolor=blue,
    citecolor=blue,
    urlcolor=blue,
    plainpages=false,
    pdfstartview=FitH,
    pdfborder={0 0 0},
    linktocpage}

\begin{document}
\title{Joint Estimation of a Two-Phase Spin Rotation beyond Classical Limit}
\author{Jiahao Cao}
\affiliation{Beijing Academy of Quantum Information Sciences, Beijing 100193, China}
\affiliation{State Key Laboratory of Low Dimensional Quantum Physics, Department of Physics, Tsinghua University, Beijing 100084, China}
\author{Xinwei Li}
\email{xinwei.li1991@outlook.com}
\affiliation{Beijing Academy of Quantum Information Sciences, Beijing 100193, China}
\affiliation{Graduate School of China Academy of Engineering Physics, Beijing 100193, P. R. China}
\author{Tianwei Mao}
\affiliation{State Key Laboratory of Low Dimensional Quantum Physics, Department of Physics, Tsinghua University, Beijing 100084, China}
\author{Wenxin Xu}
\affiliation{State Key Laboratory of Low Dimensional Quantum Physics, Department of Physics, Tsinghua University, Beijing 100084, China}
\author{Li You}
\email{lyou@tsinghua.edu.cn}
\affiliation{Beijing Academy of Quantum Information Sciences, Beijing 100193, China}
\affiliation{State Key Laboratory of Low Dimensional Quantum Physics, Department of Physics, Tsinghua University, Beijing 100084, China}
\affiliation{Frontier Science Center for Quantum Information, Beijing, China}
\affiliation{Collaborative Innovation Center of Quantum Matter, Beijing 100084, China}
\affiliation{Hefei National Laboratory, Hefei, Anhui 230088, China}

\date{\today}

\begin{abstract}
Quantum metrology employs entanglement to enhance measurement precision~\cite{giovannetti2004quantum, giovannetti2011advances, pezze2018quantum}. The focus and progress so far have primarily centered on estimating a single parameter. In diverse application scenarios, estimation of more than one single parameter is often required. Joint estimation of multiple parameters can benefit from additional advantages for further enhanced precision. Here we report quantum-enhanced estimation of simultaneous spin rotations around two orthogonal axes, making use of spin-nematic squeezing in an atomic Bose-Einstein condensate. Aided by the $F=2$ atomic ground hyperfine manifold coupled to the nematic-squeezed $F=1$ states as an auxiliary field through a sequence of microwave (MW) pulses, multiple spin-1 observables are simultaneously measured, reaching an enhancement factor 3.3 to 6.3 decibels (dB) beyond the classical limit over a wide range of rotation angles. Our work realizes the first quantum enhanced multiparameter estimation using entangled massive particles. The techniques developed and the protocols implemented also highlight the application of two-mode squeezed vacuum states in quantum-enhanced sensing of noncommuting spin rotations simultaneously.
\end{abstract}

\maketitle

Quantum-enhanced sensing improves parameter estimation precision beyond the standard quantum limit (SQL), or the classical limit~\cite{giovannetti2004quantum, giovannetti2011advances, pezze2018quantum}. The fruitful recent achievement at advanced LIGO with squeezed light~\cite{caves1981quantum}, yielding an increased number of detection events~\cite{aasi2013enhanced}, has made quantum-enhanced sensing one of the most promising quantum information technologies in the near future. Such enhanced sensing has been demonstrated in experiments with photons~\cite{polino2020photonic}, neutral atoms~\cite{appel2009mesoscopic,schleier2010states,gross2010nonlinear,riedel2010atom,lucke2011twin,sewell2012magnetic,bohnet2014reduced,hosten2016measurement,luo2017deterministic,colombo2022time}, trapped ions~\cite{gilmore2021quantum,marciniak2022optimal}, and artificial atoms~\cite{wang2019heisenberg}. Significant technological gains are expected with atomic sensors such as in clocks~\cite{pedrozo2020entanglement} and magnetometers~\cite{muessel2014scalable}.

Ideas for quantum-enhanced metrology have been extended to multiparameter settings as well, where estimation of multiple parameters can reach beyond SQL simultaneously~\cite{szczykulska2016multi,liu2020quantum}. They are relevant for a broad range of applications, such as imaging and microscopy~\cite{albarelli2020perspective,bisketzi2019quantum}, multidimensional field sensing~\cite{baumgratz2016quantum,meng2023machine}, and quantum networks for atomic clocks~\cite{komar2014quantum} etc., and they have motivated growing interests from both theoretical and experimental perspectives. A significant challenge in multiparameter estimation arises when the generators of parameter encoding do not commute. Observables that provide maximal information about different parameters need not necessarily commute~\cite{pezze2017optimal}, which could potentially impair quantum-enhanced sensitivities for all the parameters. Recently, Lorcán O. Conlon \textit{et al}. have experimentally demonstrated optimal single- and two-copy collective measurements for simultaneously estimating two noncommuting qubit rotations~\cite{conlon2023approaching}. 

Multimode squeezed states, which are often related to quantum entanglement and Einstein-Podolsky-Rosen steering, are essential resources for quantum-enhanced multiparameter estimation. The idea was first delineated in the form of quantum dense coding~\cite{braunstein2000dense} and was later demonstrated experimentally with the two-mode squeezed vacuum state generated by optical parametric amplifiers~\cite{li2002quantum,steinlechner2013quantum, liu2018loss, du2020quantum}. A related recent study~\cite{ustc2023} reports spatial displacement estimation beyond the SQL along two quadratures of its center-of-mass motion coordinate for a trapped ion. A two-mode squeezed state was recently prepared for a trapped ion motion using additional radio potentials without laser field and demonstrated SU(1,1) interferometry~\cite{Oregon2023}. 

\begin{figure*}[!htp] 
	\centering
	\includegraphics[width=1.0\linewidth]{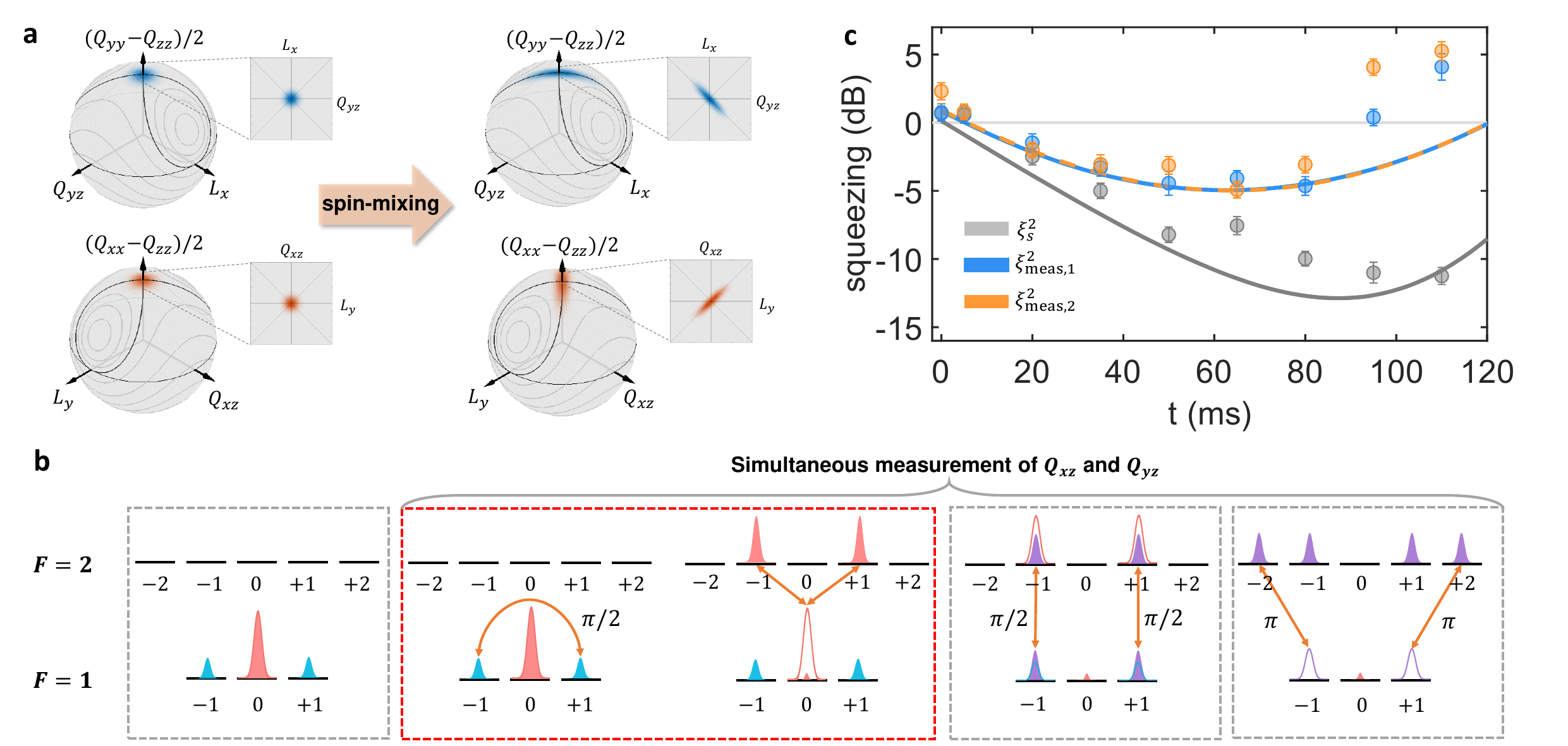}
	\caption{Spin-nematic squeezing in both SU(2) subspaces. (a) The generation of spin-nematic squeezed state is illustrated by quasiprobability distributions on Bloch spheres for both SU(2) subspaces $\left\{L_y, Q_{xz},\left(Q_{xx}-Q_{zz}\right) / 2\right\}$ and $\left\{Q_{yz}, L_x,\left(Q_{yy}-Q_{zz}\right) / 2\right\}$, simulated here for ${N = 100}$ atoms. The initial polar state corresponds to all atoms in the $m_F=0$ component. Evolving at $q\approx|c_2|$, spin-nematic squeezing quickly develops along the separatrix (dark gray lines) on both Bloch spheres. (b) Schematics illustrating the simultaneous measurement of $Q_{xz}$ and $Q_{yz}$. From left to right, the first gray dotted line box illustrates the atomic number distribution of the initial nematic-squeezed state. Well-designed MW pulses realize a $\pi/2$-pulse coupling between $a_{1,+1}$ and $a_{1,-1}$ while coherently splitting atoms from $a_{1,0}$ into $a_{2,+1}$ and $a_{2,-1}$ effectively (enclosed in a red dashed box). The complete experimental MW sequence is detailed in SM. A small number of atoms remain in the $|F=1, m_{F}=0\rangle$ state due to imperfections in the pulse sequence. The next two gray dotted line boxes show the described $\pi/2$-pulse couplings between $|F=1,m_{F}\pm1\rangle$ and $|F=2,m_{F}\pm1\rangle$, as well as the transfer of $|F=1,m_{F}=\pm1\rangle$ atoms to $|F=2,m_{F}=\pm2\rangle$. (c) The measured squeezing parameter at the optimal squeezing (for $q=|c_2|$) versus evolution time. The gray dots show the experimental results in a single SU(2) subspace by applying an RF rotation to couple sublevels in the $F=1$ manifold~\cite{hamley2012spin}. The blue and orange dots show the experimental results of simultaneously measured squeezing in both SU(2) subspaces. The simultaneously observed squeezing parameter diminishes at longer times due to imperfect MW pulses as well as other decoherence or dissipation mechanisms. All markers denote experimental results averaged over 100 runs and the corresponding solid and dashed lines denote numerical calculations based on truncated Wigner approximation. The blue and orange dashed lines account for Gaussian-distributed noise in the amplitude of signal mode ($a_{S/A}$) and a Gaussian-distributed random phase in the pump mode ($a_{2,\pm1}$), both calibrated as detailed in SM~\cite{supp}.}
	\label{fig:schematic}
\end{figure*}

Here we report quantum-enhanced simultaneous estimation of two unknown phases encoded in rotations about two orthogonal collective spin components $L_x$ and $L_y$ of a spin-1 atomic Bose-Einstein condensate (BEC). 
A complete description of the spin-1 system is provided by the SU(3) Cartesian dipole-quadrupole basis, which consists of the three components of the collective spin operators $L_i$ and the moments of the rank-2 quadrupoles or nematic tensors $Q_{i j}$ with $\{i, j\} \in\{x, y, z\}$. Detailed definitions of these operators are provided in Supplemental Material (SM)~\cite{supp}.
Through spin-mixing dynamics, an atomic analog of the two-mode squeezed vacuum state-a spin-nematic squeezed state-is generated~\cite{gross2011atomic,hamley2012spin,peise2015satisfying}, which exhibits squeezed quantum noise in both spin-nematic SU(2) subspaces spanned by $\left\{L_y, Q_{xz},\left(Q_{xx}-Q_{zz}\right) / 2\right\}$ and $\left\{Q_{yz}, L_x,\left(Q_{yy}-Q_{zz}\right) / 2\right\}$. To estimate the phases encoded into the probe state, we implement a protocol to measure two quadrupole moments $Q_{xz}$ and $Q_{yz}$ simultaneously, reminiscent of atomic homodyne detection demonstrated earlier~\cite{gross2011atomic}. 
$Q_{yz}$ and $Q_{xz}$ can be simultaneously measured because $\langle L_z \rangle = 0$ for the probe state used, which causes the two nematic operators to effectively commute, thus circumventing the limitations imposed by the Heisenberg uncertainty principle for noncommuting observables.
By taking the $F=2$ hyperfine states as the auxiliary ones and applying a sequence of MW pulses to transform $Q_{xz}$ and $Q_{yz}$ into the population differences of atomic hyperfine sublevels while preserving the squeezed noise correlation in the protocol to be detailed later, both $Q_{xz}$ and $Q_{yz}$ are measured from a single absorption image of a BEC, and the metrological gains observed for joint estimation of the two phases reach $3.3$ to $6.3$ dB beyond the SQL over a wide range of rotations explored, where the SQL $1/\sqrt{2N}$ is for estimating two parameters with an ensemble of $N$ spin-1 particles~\cite{liu2022nonlinear} as detailed in SM. 
Consistent with the standard practice of estimating a single parameter beyond the classical limit, we carry out strong (collapsing) measurements to obtain the maximal amount of information by adopting atomic homodyne detection.

\begin{figure*}[!htp]
	\centering
	\includegraphics[width=1.0\linewidth]{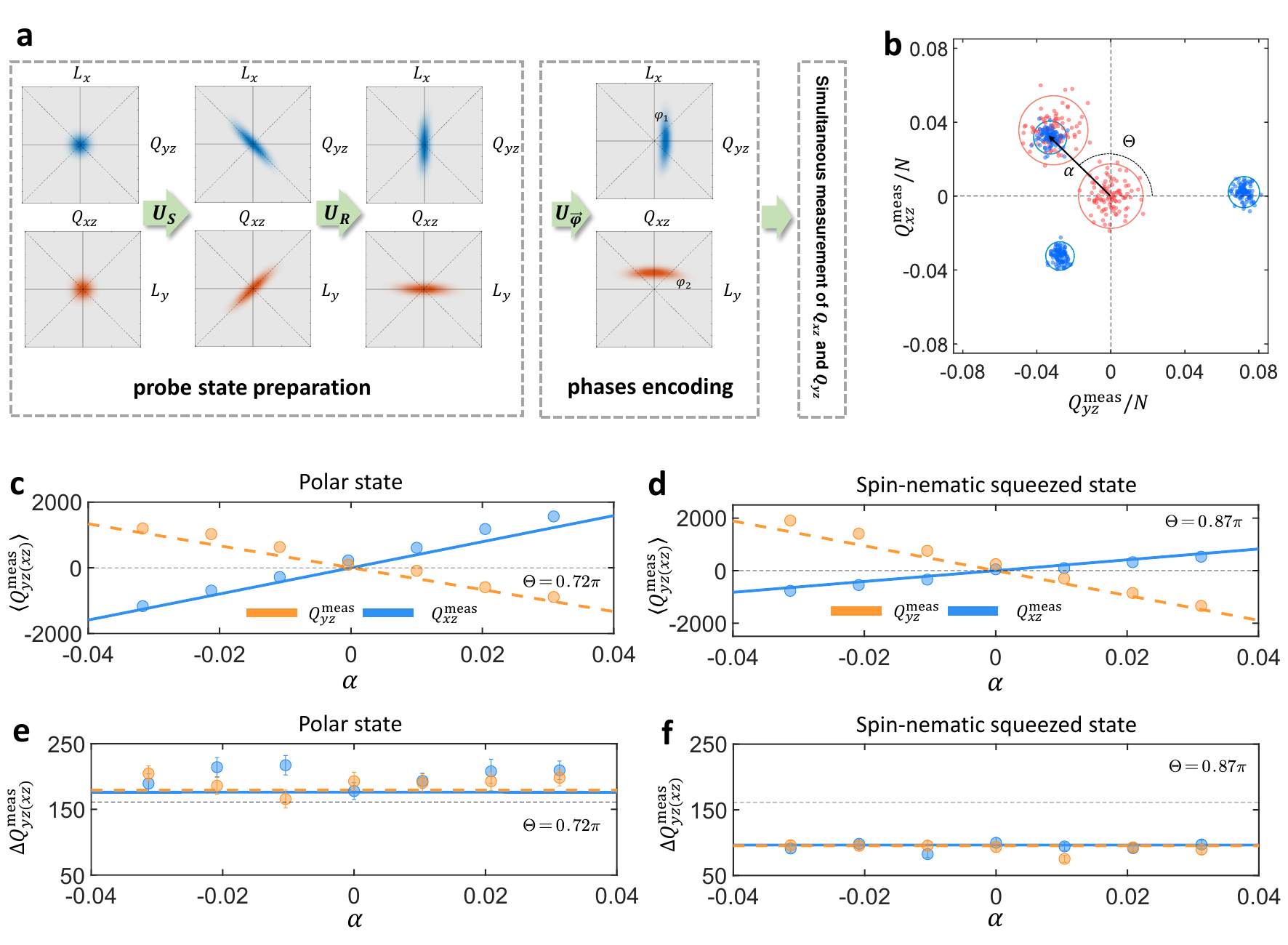}
	\caption{Joint estimation of spin rotation angles around $\hat L_x$ and $\hat L_y$. (a) Schematics of the joint estimation interferometry protocol with quasiprobability distributions for the states plotted in the $\left\{L_x, Q_{yz}\right\}$ and $\left\{L_y, Q_{xz}\right\}$ planes, respectively. $U_\mathrm{S}$, $U_\mathrm{R}$, and $U_{\vec{\varphi}}$ denote the unitary evolutions generated by spin-mixing, spinor phase rotation, and encoding spin rotation, respectively. The measurement part is shown Fig.~\ref{fig:schematic}. (b) The distributions of the simultaneously measured data for the polar state (red dots) and spin-nematic squeezed state (blue dots) sampled over 100 times after spin rotations at different $\varphi_1$ and $\varphi_2$ encoded. The distribution of polar state is much broader than that of spin-nematic squeezed state, hence the higher sensitivities for the estimation from the latter. Here $\varphi_1=\alpha \cos \Theta$ and $\varphi_2=\alpha \sin \Theta$. The red (blue) circles denote the boundaries of the measured data, $Q_{yz}^{\mathrm{meas}}$ and $Q_{xz}^{\mathrm{meas}}$, for polar (spin-nematic squeezed) state. Each circle is centered at ($\langle Q_{yz}^{\mathrm{meas}}\rangle/N$, $\langle Q_{xz}^{\mathrm{meas}}\rangle/N$), with a radius of $\sqrt{3}\sqrt{(\Delta Q_{yz}^{\mathrm{meas}})^2+(\Delta Q_{xz}^{\mathrm{meas}})^2}/N$. (c)-(f) The mean values and standard deviations of the measured $Q_{xz}$ and $Q_{yz}$ at variable $\alpha$ and fixed $\Theta$ from using polar state (c,e) and spin-nematic squeezed state (d,f) as the probes. Each point represents 100 repeated experiment average.}
	\label{fig:data}
\end{figure*}

The system we consider is a spinor atomic BEC in the $F=1$ ground hyperfine manifold. Under the single spatial mode approximation~\cite{law1998quantum}, it can be described by the Hamiltonian 
\begin{equation}\label{Ham}
H=\frac{c_2}{2N}\big[2({a}_1^{\dag}{a}_{-1}^{\dag}{a}_0{a}_0+\mathrm{h.c.})+(2{N}_0-1)(N-{N}_0)\big]-q{N}_0,
\end{equation}
with $a_{m_F}$ and $a_{m_F}^{\dagger}$ the creation and annihilation operators for condensed atoms of the $\left|F=1, m_F= \{0,\pm1\}\right\rangle$ component. This Hamiltonian contains an interaction part and a single atom part. The former includes a spin-mixing term generating paired atoms into $\left|F=1, m_F= \pm1\right\rangle$ from $\left|F=1, m_F= {0}\right\rangle$ and vice versa at a rate of $\left|c_2\right|$, and an energy shift term caused by elastic collisions. The second part, which is proportional to the atomic number $N_0$ of $|F=1,m_F=0\rangle$, represents an effective quadratic Zeeman shift (QZS). Its coefficient $q=q_B+q_{\mathrm{MW}}$ includes contributions from both the bias magnetic field ($q_B$) and a blue-detuned microwave (MW) dressing field ($q_{\mathrm{MW}}$) for fast and precise control~\cite{gerbier2006resonant}.

Prepared initially in the $\left|F=1, m_F= 0\right\rangle$ state or the polar state, nearly all atoms remain in the same state during the short-term spin-mixing dynamics. In this undepleted pump regime, the total number of atoms in $\left|F=1,m_F= \pm1\right\rangle$ is small, the operator $a_0$ ($a_0^{\dagger}$) can be replaced approximately by $\sqrt{N}$, and the Hamiltonian ~\eqref{Ham} reduces to the form of parametric down-conversion
\begin{equation}\label{Ham_approx}
H=c_2(a_1^{\dag} a_{-1}^{\dag}+\mathrm{h.c.}),
\end{equation}
when $q$ is adjusted to compensate for the energy shift of the elastic collision at $q=\left|c_2\right|$. The interaction in Eq.~(\ref{Ham_approx}) produces a spin-nematic squeezed state as shown in Fig.~\ref{fig:schematic}(a), analogous to the two-mode squeezed vacuum state~\cite{gross2011atomic, hamley2012spin,linnemann2016quantum} in the SU(1,1) interferometry. 
This state exhibits enhanced sensitivity to spin rotations parameterized by the two phases $\varphi_1$ and $\varphi_2$, specifically $e^{-i(\varphi_1L_x+\varphi_2 L_y)}$. For small rotation phases, the off-diagonal elements of the quantum Fisher information matrix become negligible, while the diagonal elements reduce to $4\langle L_{x}^2\rangle$ and \textbf{$4\langle L_{y}^2\rangle$}. When the spin-nematic state is rotated to induce squeezing along $Q_{yz(xz)}$ and antisqueezing along $L_{x(y)}$, the values of $4\langle L_{x(y)}^2\rangle$ exceed those of an unentangled polar state, thereby facilitating sensitivities beyond classical to both phases.

To jointly estimate both parameters $\varphi_1$ and $\varphi_2$, $Q_{yz}$ and $Q_{xz}$ need to be measured simultaneously, i.e., within one snapshot of absorption imaging.
In the undepleted pump regime, $N_0\sim N$, the collective spin operator $L_{x}$ ($L_y$) and the nematic operator $Q_{yz}$ ($Q_{xz}$) reduce to the quadrature operators $\sqrt{N} a_S^{\dagger}+\mathrm{h.c.}$ ($-i\sqrt{N} a_A^{\dagger}+\mathrm{h.c.}$) and $i\sqrt{N} a_S^{\dagger}+\mathrm{h.c.}$ ($\sqrt{N} a_A^{\dagger}+\mathrm{h.c.}$), approximately, here $a_S^{\dagger}=(a_1^{\dagger}+a_{-1}^{\dagger})/\sqrt{2}$ and $a_A^{\dagger}=(a_1^{\dagger}-a_{-1}^{\dagger})/\sqrt{2}$ are symmetric and asymmetric superpositions of the bosonic operators $a_{1}^{\dagger}$ and $a_{-1}^{\dagger}$. For the measurement of quadrature operators of electromagnetic fields, homodyne detection represents a successful approach and is implemented by mixing the signal field to the coherent pump field~\cite{walls2008input}. 
Analogously, the two-phase spin rotation $e^{-i(\varphi_1L_x+\varphi_2 L_y)}$ reduces to simultaneous displacements corresponding to $a_{S}$ and $a_{A}$. 
It is worth emphasizing that $Q_{yz}$ and $Q_{xz}$ are quadrature operators of the two orthogonal modes $a_{S}$ and $a_{A}$, respectively. They can be simultaneously measured beyond the classical limit, thus circumventing the Heisenberg uncertainty constraint for noncommuting observables. Therefore, our measurement scheme is different from the scenario of simulteneously measuring $L_x$ and $Q_{yz}$ reported in Ref.~\cite{kunkel2019simultaneous}, which effectively measure two conjugate quadratures in a single bosonic field $a_S$ and necessitates splitting the atomic population of the squeezed mode $a_S$ with auxiliary vacuum states via MW couplings, a process that introduces binomial atom-number fluctuations and limits the detectable squeezing to 3 dB below the SQL.
Simultaneous measurement of quantum squeezing in both $Q_{yz}$ and $Q_{xz}$ can approach the intrinsic variance of the spin-nematic squeezed state theoretically, thereby facilitating near-optimal sensitivities of $\varphi_1$ and $\varphi_2$. Detailed analysis with numerical calculations are provided in the SM~\cite{supp}.

Here in our system, atoms in $\left|F=1, m_F=0\right\rangle$ can be treated as a coherent pump field reference in measurements of the analogous quadrature operators $Q_{yz}$ and $Q_{xz}$~\cite{gross2011atomic,kunkel2019simultaneous}. To realize simultaneous measurement, a $\pi/2$-pulse coupling is applied between $a_{1}$ and $a_{-1}$ to form $a_S$ and $a_A$. Additionally, atoms in $|F=1,m_F=0\rangle$ are split into two coherent fields, through a tailored sequence of MW pulses with two orthogonal polarizations, which couples $F=1$ manifold and the ancillary states within the $F=2$ manifold. This sequence leads to the transformations $a_{1,1} \rightarrow a_S$, $a_{1,-1} \rightarrow a_A$, and $a_{2, \pm 1} \rightarrow a_0 / \sqrt{2}$ as shown in Fig.~\ref{fig:schematic}(b). 
To avoid misunderstanding the state index used before, the full label $\{F,m_F\}$ for the component mode is adopted here instead of the shorthand indexing by $m_F$ alone in the $F=1$ manifold.
Our current setup cannot generate pure $\sigma_{+/-}$ polarized microwaves, three MW pulses with controlled $\sigma_{+/-}$ power ratios are needed to complete the above transform (see SM).

Our experiment starts with a BEC inside a bias magnetic field of 0.828 G. $N=26500\pm250$ $^{87}$Rb atoms in the $|F=1, m_F=0\rangle$ state are confined in a crossed dipole trap at $c_2=-2\pi\times3.8$ Hz. The spin-mixing dynamics are initiated by quenching $q$ from $q\sim13|c_2|$ to $q=|c_2|$ through tuning the dressing microwave power. After a short time evolution, spin-nematic squeezed state is generated and a $\pi/4$ spinor phase rotation around the $\left(Q_{yy}-Q_{zz}\right) / 2$ and $\left(Q_{xx}-Q_{zz}\right) / 2$ axis is effected by two resonant MW pulses between $|1,0\rangle$ and $|2,0\rangle$ with $\pi/4$ phase difference. The quadrature angle of optimal squeezing is then aligned to be most sensitively detected by $Q_{yz}$ and $Q_{xz}$, ensuring the probe state is most sensitive to the two-phase rotation.

As illustrated in Fig.~\ref{fig:schematic}(c), spin-nematic squeezing gradually builds up from spin mixing and the degree of squeezing can be quantified by the parameter
\begin{equation}
\xi^2=20 \log _{10}\left(\Delta Q_{yz(xz)} / \sqrt{N}\right).
\end{equation}
The gray dots show the measured squeezing parameter when only one spin-nematic subspace is detected~\cite{hamley2012spin}. The blue and orange dots illustrate quantum noise reduction in the two spin-nematic SU(2) subspaces. Their respective squeezing parameters ideally should be equal in theory due to the relation $Q_{xz} = e^{i L_{z} \pi / 2} Q_{yz} e^{-i L_{z} \pi / 2}$, as the spin-nematic state resides within the zero-magnetization subspace $L_{z}=0$, ensuring $\Delta Q_{yz(xz)}=\Delta Q_{yz(xz)}$, or the equality of their variances. However, imperfections of the synthesized MW pulses before atom number detection bring in additional noise, e.g. from the fluctuations of signal ($a_{S}$ and $a_{A}$) amplitude and pump ($a_{2,\pm1}$) phase (see SM), which together degrades the measured squeezing. For longer evolution times, spin-nematic squeezing parameters measured in both subspaces are found worse due to the breakdown of undepleted pump approximation. Nevertheless, the observed high level of squeezing for both $Q_{xz}$ and $Q_{yz}$ indirectly benchmark our capability of synthesizing the effective transformations shown in Fig.~\ref{fig:schematic}(b).

The phase encoding operation $e^{i\left(\varphi_1 L_x+\varphi_2 L_y\right)}$ takes about 100 $\mu$s to complete and is enacted by small angle RF rotations. To ensure maximum sensitivity of the probe state to the phase imprinting procedure, the rotation axis is aligned with the antisqueezed quadrature as depicted in Fig.~\ref{fig:data}(a). In the alternatively parameterized form $\varphi_1=\alpha \cos \Theta$ and $\varphi_2=\alpha \sin \Theta$ as shown in Fig.~\ref{fig:data}(b), $\alpha$ of the imprinted phase is controlled experimentally by the duration of the RF pulse while $\Theta$ by the RF phase.

\begin{figure}[!htp]
	\centering
	\includegraphics[width=1.0\linewidth]{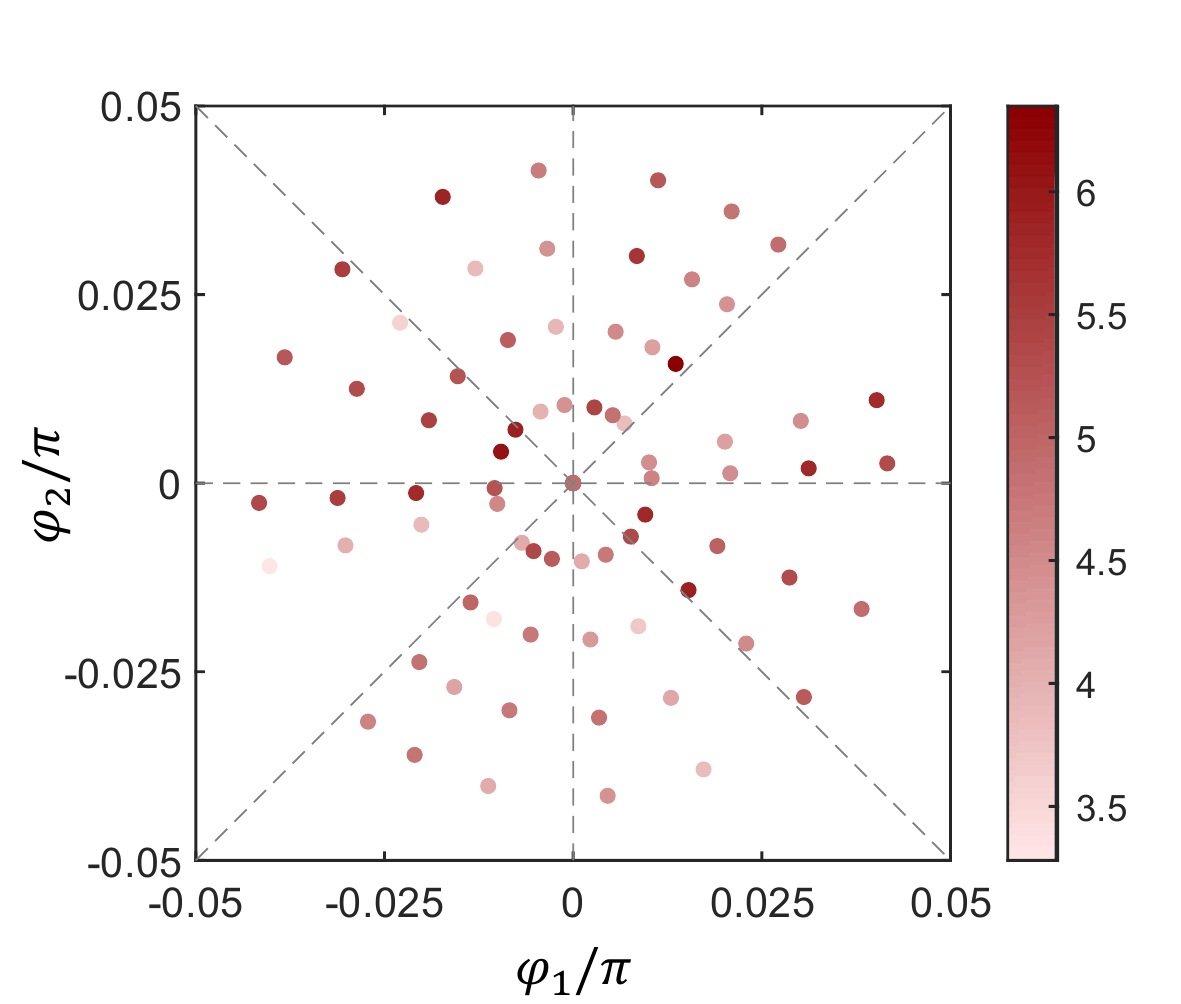}
	\caption{Metrological gains for joint estimation of two-phase rotations according to Eq.~\eqref{gain} with error propagation formulas $\Delta \varphi_1=\Delta Q_{y z}/\left|d\left\langle Q_{y z}\right\rangle/ d \varphi_1\right|$ and $\Delta \varphi_2=\Delta Q_{x z} /\left|d\left\langle Q_{x z}\right\rangle / d \varphi_2\right|$ respectively. The denominators and numerators are obtained by fitting the measured spin-nematic operators $\langle Q_{xz(yz)}^{\mathrm{meas}}\rangle$ and $\Delta Q_{xz(yz)}^{\mathrm{meas}}$ using quartic functions without linear terms, respectively. Gains of $3.3$ dB to $6.3$ dB over the SQL are observed for $\varphi_{1,2} \in[-0.05,0.05] \pi$.}
	\label{fig:metro_gain}
\end{figure}

Figures~\ref{fig:data}(c)-(f) show the parameter-dependent mean values and standard deviations of the measured $Q_{yz}$ and $Q_{xz}$ using the nonentangled polar state and the spin-nematic squeezed state as probes. For the polar state, the standard deviation observed is found to be slightly larger than the shot noise as a result of technical imperfections. For the spin-nematic squeezed state, the distributions of measured data are squeezed along both the $Q_{xz}$ and $Q_{yz}$ axes. Their mean values are linearly dependent on the parameter $\alpha$ for fixed $\Theta$, while the standard deviations are almost independent of the parameters $\alpha$ and $\Theta$. The measurement MW pulses are generated by the same AWG as the phase encoding RF pulses to ensure nearly perfect phase correlation between the two steps. The total measurement protocol takes about 374 $\mu$s to complete, during which the less-than-ideal differential phase shift between the two MW signals starts to contribute to phase noise, which altogether limits the measured squeezing and therefore the achieved metrological gain. 

To further elucidate the performance of the joint estimation protocol, in Fig.~\ref{fig:metro_gain}, the metrological gain for the two-phase estimation, defined as
\begin{eqnarray}
\zeta=-10 \log _{10}\left[\frac{\left(\Delta \varphi_1\right)^2+\left(\Delta \varphi_2\right)^2}{\left(\Delta \varphi_1\right)_{\mathrm{SQL}}^2+\left(\Delta \varphi_2\right)_{\mathrm{SQL}}^2}\right],
\label{gain}
\end{eqnarray}
is shown in the rotation range of $\varphi_{1,2} \in[-0.05,0.05] \pi$, enhancement of $3.3$ to $6.3$ dB beyond the SQL of 26 000 atoms is observed, despite of the loss of about 500 atoms during the experiment cycle. We have also conducted an analysis of the metrological gain for the polar state and a sensitivity approximately 2 dB worse than the SQL is found, due to the crosstalk noise (see SM). While the metrological gain is expected to be symmetric with respect to $\varphi_1$ and $\varphi_2$, the experimental results are affected by pump phase noise (see SM) and degrade stochastically in the $\varphi_1$-$\varphi_2$ plane.

In conclusion, we propose and experimentally demonstrate, for the first time, the simultaneous estimation of two phases generated by two orthogonal spin rotations using an entangled atomic BEC. Coherent coupling of the $F=1$ to the $F=2$ manifold, with the latter serving as an auxiliary field, enables simultaneous measurement of two observables, from which two parameters are estimated. Sensitivities exceeding 3 dB beyond the SQL are observed across a wide range of rotation angles. The metrological advantage of spin-nematic squeezed state was previously demonstrated for single-parameter estimation scenarios~\cite{mao2023quantum}. The present work expands such quantum-enhanced sensing applications to multiparameter metrology and will likely stimulate the development of a broad range of applications in atomic sensors. Limited by the coherence time of the spin-nematic squeezed state in our experiment, interaction-based readout, which recently enabled significantly enhanced metrological gains over direct linear read-out adopted here~\cite{mao2023quantum}, remains to be implemented. An operational protocol is known~\cite{li2021quantum} and its implementation in the future will mitigate a large part of the atom number detection noise after coherence time is improved.

We thank Drs Q. Liu and L. N. Wu for helpful discussions. This work is supported by the National Natural Science Foundation of China (NSFC) (Grants No. 92265205 and No. 12361131576), and by the Innovation Program for Quantum Science and Technology (2021ZD0302100).

J. C. and X. L. contributed equally to this work.


%

\clearpage

\onecolumngrid
\begin{center}
\vspace{1.5 cm}
\textbf{\large Supplemental Material for: Joint estimation of a two-phase spin rotation beyond classical limit}
\end{center}

\setcounter{figure}{0}
\setcounter{equation}{0}
\setcounter{table}{0}
\makeatletter
\renewcommand{\thefigure}{S\arabic{figure}}
\renewcommand{\theequation}{S\arabic{equation}}
\renewcommand{\thetable}{S\arabic{table}}

\section{Spin-nematic squeezing in spin-1 BEC}
\subsection{Dynamical simulations}\label{1A}
For a spin-1 Bose-Einstein condensate (BEC) composed of $N$ ${}^{87}$Rb atoms, the full spin-mixing dynamics under single spatial mode approximation (SMA)~\cite{law1998quantum} is described by the following Hamiltonian
\begin{equation}\label{SM:ham}
\begin{aligned}
H=& \frac{c_{2}}{2N}\bigl[2(a_{1}^{\dagger}a_{-1}^{\dagger}a_{0}a_{0}+a_{1}a_{-1}a_{0}^{\dagger}a_{0}^{\dagger})+(N_{1}-N_{-1})^{2}+(2N_{0}-1)(N_{1}+N_{-1})\big]-qN_{0},
\end{aligned}	
\end{equation}
where $c_2<0$ is the ferromagnetic spin-exchange interaction coefficient, and $q$ is the quadratic Zeeman shift (QZS). $a_m$ ($a_m^{\dagger}$) is the annihilation (creation) operator for atoms in the $|F=1,m=\{0,\pm1\}\rangle$ spin component and $N_m=a_m^{\dagger}a_m$ represents the associated atom number. To realistically simulate the generation of the spin-nematic squeezing in experiments, it is necessary to take into account atom loss, and the spin-mixing dynamics can then be modeled by the master equation
\begin{equation}\label{SM:master}
\frac{d \rho}{d t}=-i[H(t), \rho]+\sum_{m=0, \pm 1} \gamma_m \mathcal{D}\left[a_m\right] \rho,
\end{equation}
where $\mathcal{D}[a_{m}]\rho=a_{m}\rho a_{m}^{\dagger}-\{a_{m}^{\dagger}a_{m},\rho\}/2$ accounts for atom loss with dissipative rate $\gamma_m$ for the $|F=1,m\rangle$ component. Here we assume $\gamma_m=\gamma$ for $m=0,\pm1$, and $H(t)$ is a time-dependent Hamiltonian obtained after replacing $N$ and $c_2$ in Eq.~\eqref{SM:ham} by $N(t)=Ne^{-\gamma t}$ and $c_{2}(t)=c_{2}e^{-\gamma_{c}t}$, where $\gamma_{c}$ is the measured decay rate of $c_2$.

The above master equation can be numerically simulated with the semiclassical truncated Wigner method (TWM)~\cite{gardiner2004quantum,blakie2008dynamics}, which maps the field operators $a_m$ and $a_m^{\dagger}$ into complex numbers $\psi_{m}$ and $\psi_{m}^{*}$ with the time evolution of the system unraveled into a set of stochastic differential equations (SDEs) as given in the following
\begin{eqnarray}\label{SM:sde}
&& d\psi_1 = -ic_2'\left[\psi_0^2\psi_{-1}^*+(|\psi_1|^2-|\psi_{-1}|^2+|\psi_0|^2)\psi_1\right]dt-\frac{\gamma}{2}\psi_1dt+\sqrt{\frac{\gamma}{2}}\space d\xi_1(t), \nonumber\\
&& d\psi_0 = -ic_2'\left[2\psi_1\psi_{-1}\psi_0^*+(|\psi_1|^2+|\psi_{-1}|^2)\psi_0\right]dt+iq\psi_0dt-\frac{\gamma}{2}\psi_0dt+\sqrt{\frac{\gamma}{2}}\space d\xi_0(t),\\
&& d\psi_{-1} = -ic_2'\left[\psi_0^2\psi_1^*+(|\psi_{-1}|^2-|\psi_1|^2+|\psi_0|^2)\psi_{-1}\right]dt-\frac{\gamma}{2}\psi_{-1}dt+\sqrt{\frac{\gamma}{2}}\space d\xi_{-1}(t).  \nonumber
\end{eqnarray}

In the above, $c_2^{\prime}=c_2(t)/N(t)$ denotes the drifting spin-exchange rate in the presence of atom loss. $d\xi_m(t)$ represents a complex Wiener noise increment satisfying $\overline{d\xi_m(t)}=0$ and $\overline{d\xi_m^*(t)d\xi_n(t)}=\delta_{m,n}dt$. The above SDEs can describe quantum evolution of the Master Eq.~\eqref{SM:master} after averaged over an ensemble of trajectories. The effect of quantum noise is incorporated into the probability distribution of the initial polar state, which is given by the Wigner function and can be sampled according to
\begin{equation}
\begin{aligned}
&\psi_{\pm1}=\dfrac{1}{2}(\alpha_{\pm1}+i\beta_{\pm1}),\\
&\psi_0=\sqrt{N}+\dfrac{1}{2}(\alpha_{\pm0}+i\beta_{\pm0}),
\end{aligned}
\end{equation}
where $\alpha_m$ and $\beta_m$ are independent real numbers following a standard normal distribution. The averages and variances of mode populations are then obtained from the averages of simulated trajectories, calculated according to $\langle{\hat N}_m\rangle=\overline{\psi_m^*\psi_m}-1/2$ and $(\Delta \hat {N}_m)^2=(\Delta\psi_m^*\psi_m)^2-1/4$.  We take 10000 trajectories for most of this study.

The parameters used in our simulations include initial atom number $N=26000$, $c_2=-2\pi\times3.8$ Hz, $q=2\pi\times3.9$ Hz, and loss rate $\gamma=0.069~\rm{s^{-1}}$, all of which are independently calibrated in experiments. We also experimentally compensate for the influence of atom loss by ramping $q$ with a rate $\gamma_q=\gamma_c$ to keep $c_2(t)/q(t)$ fixed as in Refs.~\cite{liu2022nonlinear,mao2023quantum}.

\subsection{SU(3) operators and Bloch spheres}\label{1B}
To better understand the spin-mixing dynamics in phase spaces, both the angular momentum $L_i$ and the nematic tensor $Q_{i,j}$ ($i,j\in x,y,z$) operators, which constitute SU(3) Lie algebra, are needed. The definition of the nematic tensor operators are as follows
\begin{equation}\label{SM:nematic}
\begin{aligned}
&Q_{yz}=\frac{i}{\sqrt{2}}\left(-a_1^{\dagger} a_0+a_0^{\dagger} a_1-a_{-1}^{\dagger} a_0+a_0^{\dagger} a_{-1}\right),\\
&Q_{xz}=\frac{1}{\sqrt{2}}\left(a_1^{\dagger}a_0+a_0^{\dagger}a_1-a_{-1}^{\dagger} a_0-a_0^{\dagger} a_{-1}\right),\\
&Q_{xx}=\frac{2}{3}a_0^{\dagger}a_0-\frac{1}{3}a_1^{\dagger}a_{1}-\frac{1}{3}a_{-1}^{\dagger}a_{-1}+a_{-1}^{\dagger}a_1+a_1^{\dagger}a_{-1},\\
&Q_{yy}=\frac{2}{3}a_0^{\dagger}a_0-\frac{1}{3}a_1^{\dagger}a_{1}-\frac{1}{3}a_{-1}^{\dagger}a_{-1}-\hat{a}_{-1}^{\dagger}a_1-a_1^{\dagger}a_{-1},\\
&Q_{zz}=\frac{2}{3}a_1^{\dagger}a_1-\frac{4}{3}a_0^{\dagger}a_0+\frac{2}{3}a_{-1}^{\dagger}a_{-1},
\end{aligned}
\end{equation}
and the spin-1 angular momentum operators are
\begin{equation}\label{SM:angular}
\begin{aligned}
&L_x=\frac{1}{\sqrt{2}}\left(a_1^{\dagger}a_0+a_0^{\dagger}a_1+a_{-1}^{\dagger}a_0+a_0^{\dagger} a_{-1}\right), \\
&L_y=\frac{i}{\sqrt{2}}\left(-a_1^{\dagger}a_0-a_0^{\dagger}a_{-1}+a_0^{\dagger} a_1+a_{-1}^{\dagger} a_0\right), \\
&L_z=a_1^{\dagger}a_1-a_{-1}^{\dagger}a_{-1}.
\end{aligned}
\end{equation}

The above eight operators are linearly independent, and they form seven triple operator combinations that span SU(2) subspaces. The spin-nematic squeezing can be illustrated in the subspace Bloch spheres formed by $\{Q_{yz},L_{x},(Q_{yy}-Q_{zz})/2\}$ and $\{L_{y},Q_{xz},(Q_{xx}-Q_{zz})/2\}$ as shown in the main text.

\subsection{Validity of the single-mode approximation in our experiment}\label{1C}
In the experiment, we prepare a condensate of $N=26500\pm250$ ${}^{87}$Rb atoms, tightly confined within an optical dipole trap with frequencies $\omega_{x,y,z}=2\pi\times\left(118,92,185\right)$Hz. The s-wave scattering lengths for the total spin-0 and spin-2 channels are $a_0=101.8a_B$ and $a_2=100.4a_B$ ($a_B$ denotes the Bohr radius), respectively. Consequently, the Thomas-Fermi radii of the condensate are $R_x=4.8 \mu$m, $R_y=6.1 \mu$m and $R_z=3.0 \mu$m. These parameters yield a spin healing length of approximately $25 \mu$m, which is much larger than the size of the condensate. Furthermore, the magnetic field gradient in our setup is sufficiently small, ensuring that the overall trapping potential is the same for all three spin components. Given these considerations, the condensate can be well described by the single-mode approximation.

\section{Simultaneous measurements of $Q_{xz}$ and $Q_{yz}$}
\subsection{Quadrature squeezing and atomic homodyne detection}\label{2A}
For the short-time evolution starting from polar state, atom number in the $|F=1,m=0\rangle$ component is approximately undepleted. At $q=|c_2|$, the complete Hamiltonian approximately takes the two-mode squeezing form after the replacement of $\hat{a}_{0}^{\dagger}\rightarrow\sqrt{N}e^{i\phi_{0}}$
\begin{equation}\label{SM:ham_undep}
H_{\mathrm{udp}}=c_2(e^{2i\phi_{0}}a_1 a_{-1}+e^{-2i\phi_{0}}a_1^\dagger a_{-1}^\dagger),
\end{equation}
with $\phi_0=0$. Equation~\eqref{SM:ham_undep} thus gives the equivalent two-mode squeezing dynamics as in undepleted pump optical four-wave mixing (FWM) or parametric down conversion, and the relevant angular momentum operators $L_x$ and $L_y$ as well as the nematic tensor operators $Q_{yz}$ and $Q_{xz}$ all reduce to quadrature operators
\begin{equation}\label{SM:quadrature}
\begin{aligned}
&L_{x}=\sqrt{\frac{N}{2}}[(a_{+1}^{\dagger}+a_{-1}^{\dagger})+(a_{+1}+a_{-1})],\\
&L_{y}=i\sqrt{\frac{N}{2}}[(a_{-1}^{\dagger}-a_{+1}^{\dagger})+(a_{+1}-a_{-1})],\\
&Q_{yz}=-i\sqrt{\frac{N}{2}}[(a_{+1}^{\dagger}+a_{-1}^{\dagger})-(a_{+1}+a_{-1})],\\
&Q_{xz}=\sqrt{\frac{N}{2}}[(a_{+1}^{\dagger}-a_{-1}^{\dagger})+(a_{+1}-a_{-1})].
\end{aligned}
\end{equation}
It is then straightforward to see that $Q_{yz}$ and $Q_{xz}$ can be measured following atomic homodyne detection~\cite{gross2011atomic} with two coherent pumps of a relative $\pi/2$ phase difference. 

In the simultaneous measurement scheme adopted for our experiment as depicted in Fig. 1(b) of the main text, the operations in the red dashed box execute the transformation of 
\begin{equation}\label{SM:transf}
\begin{aligned}
a_{1,-1}^{\prime}&=\frac{1}{\sqrt{2}}a_{1,+1}+\frac{1}{\sqrt{2}}a_{1,-1}=a_{S},\\
a_{1,+1}^{\prime}&=\frac{1}{\sqrt{2}}a_{1,+1}-\frac{1}{\sqrt{2}}a_{1,-1}=a_{A},
\end{aligned}
\end{equation}
and split atoms in the $|F=1,m=0\rangle$ component into two halves
\begin{equation}\label{SM:pump}
\begin{aligned}
a_{2,-1}^{\prime}&=i\varepsilon_1 a_{1,0},\\
a_{2,+1}^{\prime}&=\varepsilon_2 a_{1,0}.
\end{aligned}
\end{equation}
Here for simplicity, we take $\varepsilon_1 \approx \varepsilon_2 \approx 1/\sqrt{2}$, while in realistic experiments a small number of atoms (less than $1000$) could remain in the $|F=1,m=0\rangle$ component, which can be neglected as the population of the pump is relatively large. Analogous to optical homodyne detection, $a_{1,\pm1}$, or equivalently $a_{S(A)}$, is the signal field that contains the squeezing. While $a_{2,\pm1}$ is the coherent pump, often referred to as the ``local oscillator" in the context of optical homodyne detection.

The next two step coupling the signals $a_S$ and $a_A$ to the coherent pump enables atomic homodyne detection simultaneously. After the microwave (MW) pulses are applied, we obtain
\begin{equation}\label{SM:trans_final}
\begin{aligned}
a_{2,-2}^{\mathrm{out}}&=\frac{1}{\sqrt{2}}\left(\frac{1}{\sqrt{2}}a_{1,+1}+\frac{1}{\sqrt{2}}a_{1,-1}+i\varepsilon_{1}a_{1,0}\right), \\
a_{2,-1}^{\mathrm{out}}&=\frac{1}{\sqrt{2}}\left(\frac{i}{\sqrt{2}}a_{1,+1}+\frac{i}{\sqrt{2}}a_{1,-1}+\varepsilon_{1}a_{1,0}\right), \\
a_{2,+2}^{\mathrm{out}}&=\frac{1}{\sqrt{2}}\left(\frac{1}{\sqrt{2}}a_{1,+1}-\frac{1}{\sqrt{2}}a_{1,-1}-\varepsilon_{2}a_{1,0}\right), \\
a_{2,+1}^{\mathrm{out}}&=\frac{1}{\sqrt{2}}\left(\frac{1}{\sqrt{2}}a_{1,+1}-\frac{1}{\sqrt{2}}a_{1,-1}+\varepsilon_{2}a_{1,0}\right),
\end{aligned}
\end{equation}
with the final step transfering atoms in $|F=1,m=1\rangle$ and $|F=1,m=-1\rangle$ into $|F=2,m=1\rangle$ and $|F=2,m=1\rangle$ components respectively. 
The populations in the above $F=2$ 
components can be detected simultaneously from absorption imaging of the $F=1$ components. Using Eq.~\eqref{SM:trans_final}, we obtain $Q_{xz}$ and $Q_{yz}$ in a single experimental run according to
\begin{equation}\small
\begin{aligned}
N_{2,-1}^{\mathrm{out}}-N_{1,-1}^{\mathrm{out}}=\frac{i{\varepsilon}_{1}}{\sqrt{2}}\big(a_{1,0}^{\dagger}a_{1,+1}+a_{1,0}^{\dagger}a_{1,-1}\big)+\mathrm{H.c.}=\varepsilon_{1}Q_{yz},\\
N_{2,+1}^{\mathrm{out}}-N_{1,+1}^{\mathrm{out}}=\frac{\varepsilon_{2}}{\sqrt{2}}\big(a_{1,+1}^{\dagger}a_{1,0}-a_{1,-1}^{\dagger}a_{1,0}\big)+\mathrm{H.c.}=\varepsilon_{2}Q_{xz}.
\end{aligned}
\end{equation}
\subsection{Experimental MW pulse sequence}
\begin{figure*}[!htp]
  \centering
  \includegraphics[width=1.0\linewidth]{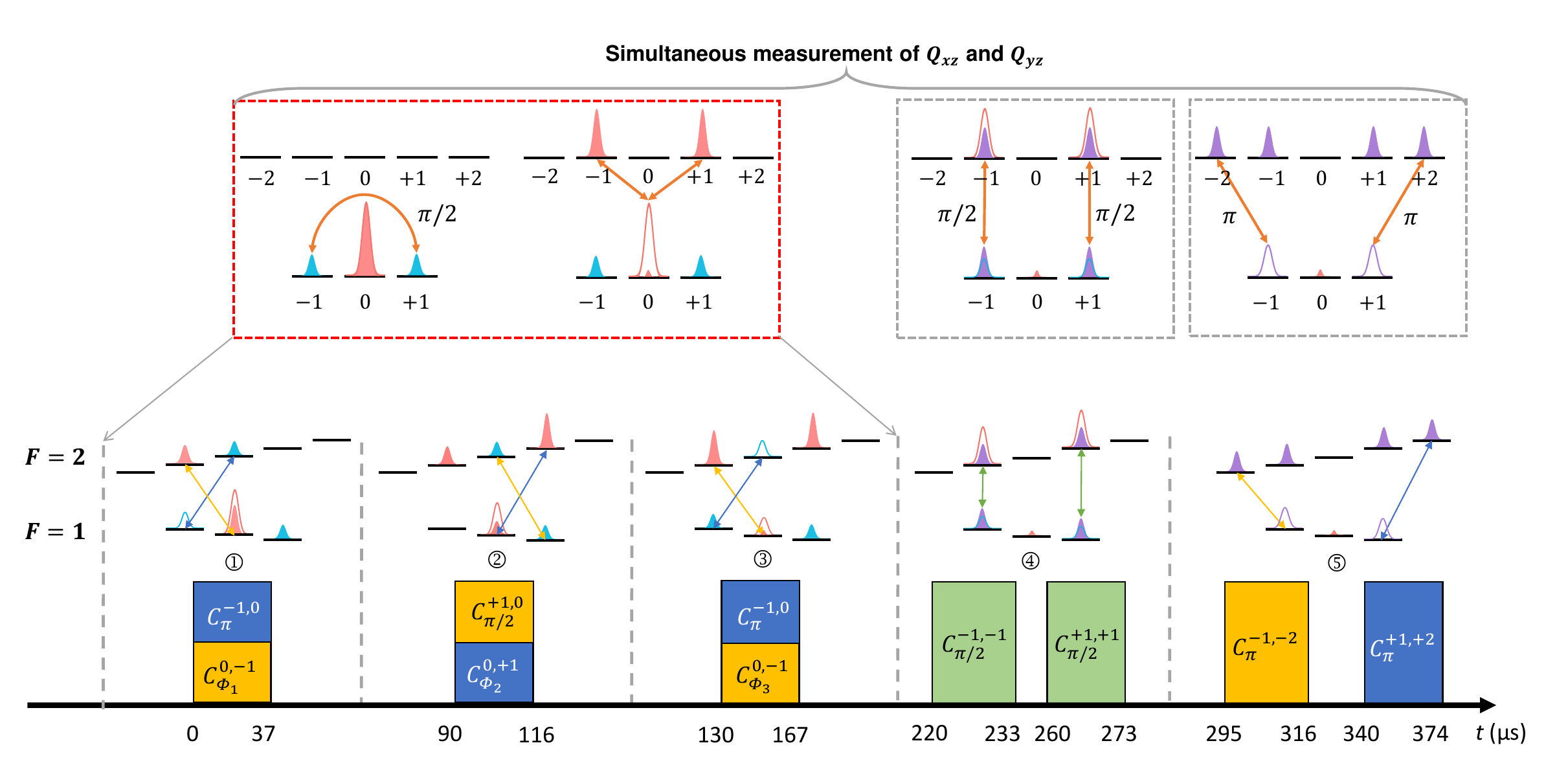}
  \caption{The actual MW pulse sequence for simultaneous measurement of $Q_{yz}$ and $Q_{xz}$.}
  \label{fig:pulse}
\end{figure*}

Limited by our experimental setup, the actual MW pulses applied for simultaneous measurement of $Q_{yz}$ and $Q_{xz}$ experimentally is shown in Fig.~\ref{fig:pulse}. The lower panel shows the MW pulse measurement sequence and the upper panel illustrates atom number distributions in all sublevels before (empty shape) and after (filled shape) the MW pulse sequence is implemented. The complete measurement protocol lasts for about 374 $\mu$s over a single experiment run. The blue, yellow, and green squares correspond to MW pulses with $\sigma_{+}$, $\sigma_{-}$, and $\pi-$polarizations, respectively. The blue and red shapes denote the signals and pump atomic modes for atomic homodyne detection, while the purple shapes depict the atoms after $\pi/2$-pulse couplings between the signals and the pumps. A small number of atoms remain in the $|F=1,m_{F}=0\rangle$ state as a result of  imperfect pulse sequence. Here $C^{i,j}_{\varPhi}$ denotes the coherent coupling between $\left|F=1, m_F=i\right\rangle$ and $\left|F=2, m_F=j\right\rangle$ at a rotation angle $\varPhi$. For the first three pulses, $\varPhi_1$,$\varPhi_2$, and $\varPhi_3$ are adjusted by changing the power ratio between $\sigma_{+}$ and $\sigma_{-}$ polarizations to approximately split the atoms in $\left|F=1, m_F=0\right\rangle$ into two equal halves coherently.

Next, we describe the pulses in detail below. The $\pi/2$ coherent coupling between (1, ±1) and the coherent splitting of (1, 0) into (2, ±1) are achieved by the first three microwave pulses with only a small number of atoms (less than 1000) remaining in (1,0), using $F=2$ manifold as ancillary.
\begin{enumerate}
\item The first pulse facilitates a $\pi$-pulse coupling between (1, -1) and (2, 0), transferring atoms from (1, -1) to (2,0). Given that the microwave resonant with the transition between (1, -1) and (2,0) is also in resonance with the transition between (1, 0) and (2, -1), albeit with different polarizations, a part of atoms in (1, 0) is concurrently transferred to state (2, -1) as well.

\item The second microwave pulse induces a $\pi/2$-pulse coupling between (2, 0) and (1, 1), which forms $a_{S}=(a_{1}+a_{-1})/\sqrt{2}$ and $a_{A}=(a_{1}-a_{-1})/\sqrt{2}$. Since the microwave frequency resonant with the transition from (2, 0) to (1, 1) also matches the resonance condition for the transition between (1, 0) and (2, 1), a part of the atomic population in (1, 0) is transferred to (2, 1) as a result of this pulse.

\item The third microwave pulse replicates the $\pi$-pulse coupling between (1, -1) and (2, 0), and also transfers a part of the atoms in (1, 0) to (2, -1). 

\item The fourth and fifth microwave pulses induce $\pi/2$-pulse couplings between (2, ±1) and (1, ±1), which enable the homodyne couplings between $a_{S(A)}$ and the coherent pump. 

\item The sixth and seventh microwave pulses facilitate $\pi$-pulse couplings between (1, ±1) and (2, ±2), transferring atoms from (1, ±1) to (2, ±2) for subsequent absorption imaging.
\end{enumerate}
The above microwave pulses are synthesized by signals radiated from two different antennas, whose $\sigma_{+/-}$ polarization ratios are different from each other. By tuning the relative phase and the amplitude of the two signals, we can adjust the polarization of the synthesized pulse. However, it is important to note that our microwave generators operate optimally within a restricted power range, which complicates the generation of pulses with a pure $\sigma_{+}$ or $\sigma_{-}$ polarization at the desired amplitude. 

\begin{figure}[!htp]
\centering
\includegraphics[width=0.75\linewidth]{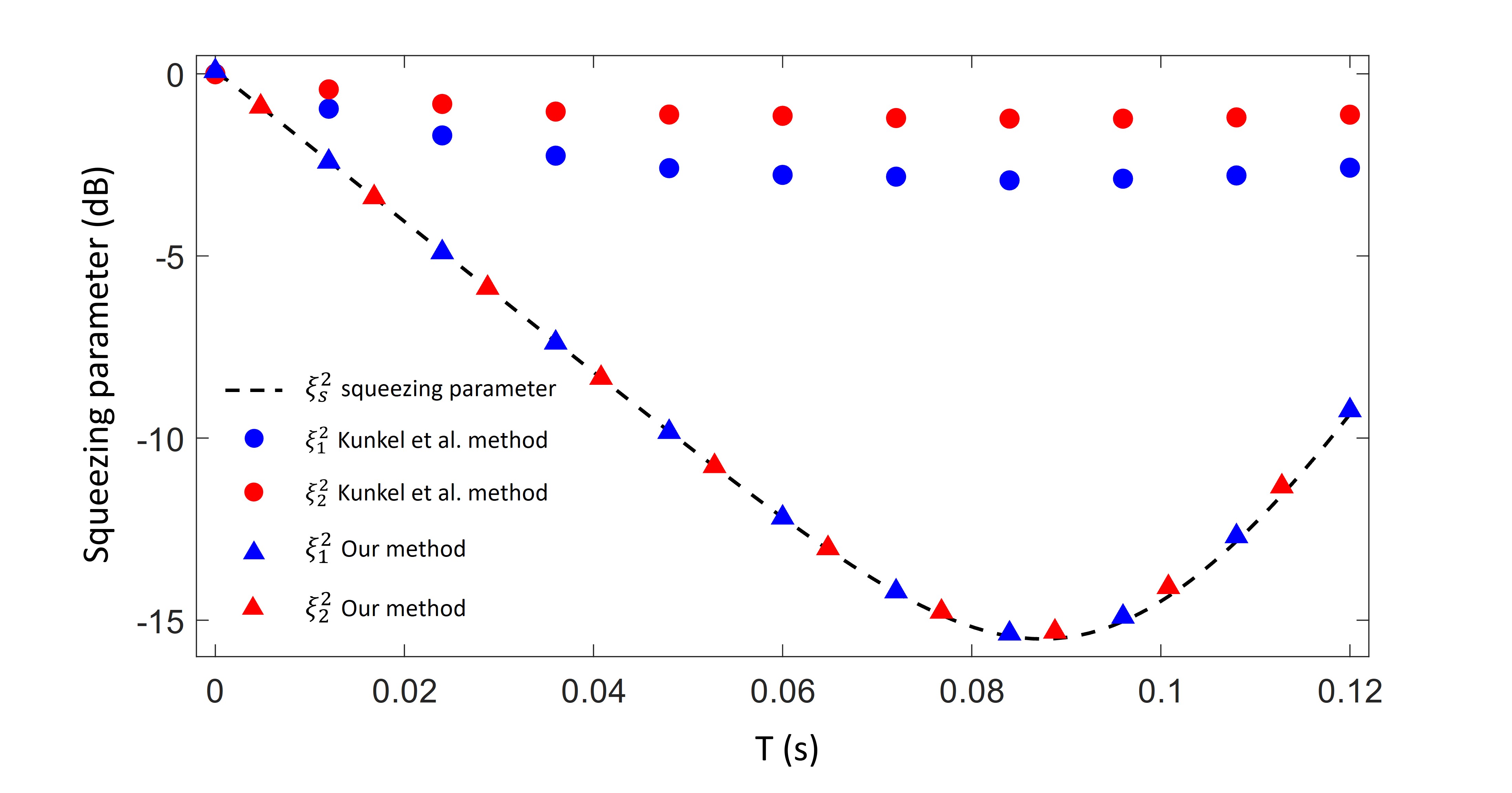}
\caption{The numerical simulation of squeezing parameter measured simultaneously in both subspaces (without considering technical noises). The black dashed line is the inherent squeezing parameter of the spin-nematic squeezed state. The blue and red triangles correspond to the numerical simulation of our method for measuring squeezing in both subspaces. The blue and red circles correspond to the numerical simulation using the method in Ref.~\cite{kunkel2019simultaneous}. The parameters used for numerical simulation are identical to those in our manuscript.}
\label{fig:compare}
\end{figure}

Similar to the method in Ref.~\cite{kunkel2019simultaneous}, our measurements utilize the ancillary $F=2$. However, our implemented measurement pulse sequence and the results we obtain significantly differ from the earlier work. The method reported in Ref.~\cite{kunkel2019simultaneous} is restricted to achieving detectable squeezing of up to 3 dB beyond the standard quantum limit (SQL). The partial degradation of detected squeezing is attributed to the transfer of half of the atoms from in $|F=1,m_{F}=\pm1\rangle$ to the initially empty $|F=2,m_{F}=\pm1\rangle$. Such a process mixes the squeezed mode $a_{1,\pm1}$ with the vacuum mode, consequently amplifying the quantum noise of the squeezed mode. In contrast, our developed measurement sequence is theoretically capable of detecting quantum squeezing close to the quantum state's intrinsic variance of the observables, thereby achieving near-optimal sensitivities. This enhanced performance is due to the ability of our designed sequence, as depicted in Fig.~\ref{fig:pulse}, to effectively prevent the mixing of squeezing modes with the vacuum noise. In Fig.~\ref{fig:compare}, we show the numerical simulations of measuring spin-nematic squeezing in both SU(2) subspaces using the two methods, without considering other technical noises. The comparison between the two methods is obvious. The black dashed line represents the squeezing parameter of the quantum state as system evolves over time. The blue and red triangles illustrate the numerical simulation of our method for measuring squeezing in both subspaces, which approximates the squeezing of the spin-nematic squeezed state generated in our system. The blue and red circles correspond to the numerical simulation of the squeezing parameter using method in Ref.~\cite{kunkel2019simultaneous}, demonstrating that the maximum achievable squeezing is limited to 3 dB. 

\subsection{Influence of crosstalk between ``pump" and ``signal"}\label{2B}
The first three steps of the measurement scheme split the pump mode $a_{1,0}$ into two halves and establish a $\pi/2$ SU(2) rotation between $a_{1,1}$ and $a_{1,-1}$. In the ideal situation, the signals are not affected by pumps as their polarizations are orthogonal. Unwanted couplings are, however, inevitable between sublevels of $F=1$ and $F=2$, and they introduce noise into the signal modes. Such imperfections impair the simultaneously measured squeezing in both subspaces, whose influence can be simulated by adding Gaussian distributed noise to the signal modes according to
\begin{equation}\label{SM:crosstalk_noise}
\begin{aligned}
a_S=\frac{1}{\sqrt{2}} a_{1,+1}+\frac{1}{\sqrt{2}} a_{1,-1}+\eta_S, \\
a_A=\frac{1}{\sqrt{2}} a_{1,+1}-\frac{1}{\sqrt{2}} a_{1,-1}+\eta_A,
\end{aligned}
\end{equation}
where $\eta_S$ and $\eta_A$ represent complex sources of white noise satisfying $\left\langle\eta_S\right\rangle=\left\langle\eta_A\right\rangle=0$. The noise strength for $a_S$ and $a_A$ are assumed to be equal $\delta \eta_S=\delta \eta_A=\delta \eta$ for simplicity here.

\begin{figure}[!htp]
\centering
\includegraphics[width=0.75\linewidth]{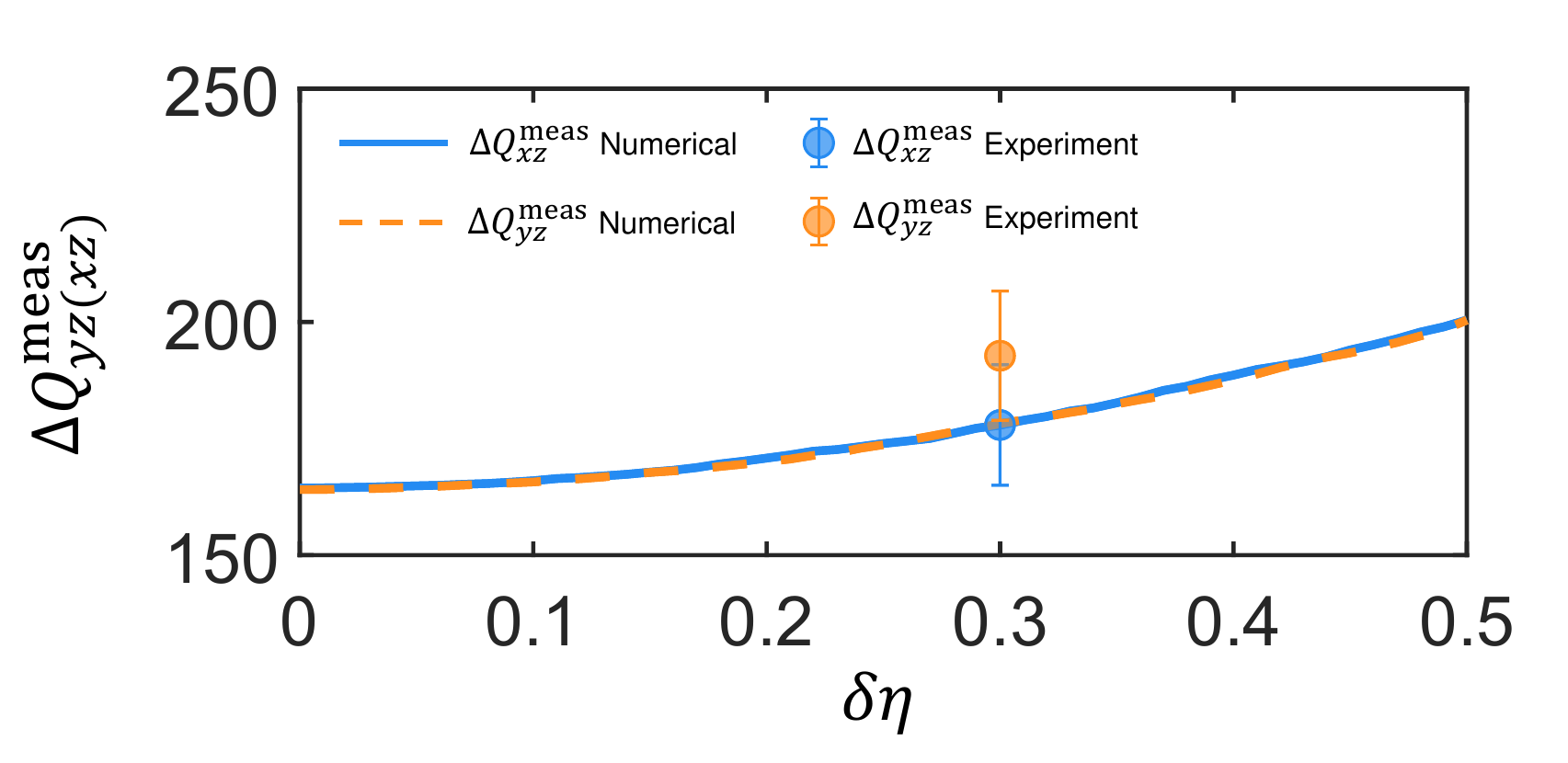}
\caption{Calibration of the signal-pump crosstalk noise strength with polar state. The blue solid line represents the numerical simulations of $\Delta Q_{xz}^{\mathrm{meas}}$ as a function of the crosstalk noise strength, while the orange color dashed line corresponds to that of  $\Delta Q_{yz}^{\mathrm{meas}}$. The experimental results are depicted by the blue and orange dots, which align reasonably close to the corresponding numerical simulations at $\delta \eta \approx 0.3$.}
\label{fig:cross_calib}
\end{figure}

To calibrate the strength of the noise $\delta \eta$, we perform numerical calculations to assess the fluctuations of the simultaneously measured $Q_{yz}$ and $Q_{xz}$ for polar state, which remains unaffected by pump phase noise. Figure~\ref{fig:cross_calib} illustrates a comparison between the experimental results and our numerical simulations, which pins down a value of approximately $\delta \eta = 0.3$. It is worth pointing out that our numerical simulations also incorporate a calibrated detection noise of ${\Delta(N_{2,\pm1}^\text{out}-N_{1,\pm1}^{\text{out}}})=24$ in the atomic population difference.

\subsection{Influence of pump mode phase noise}\label{2C}
\begin{figure}[!htp]
\includegraphics[width=0.75\linewidth]{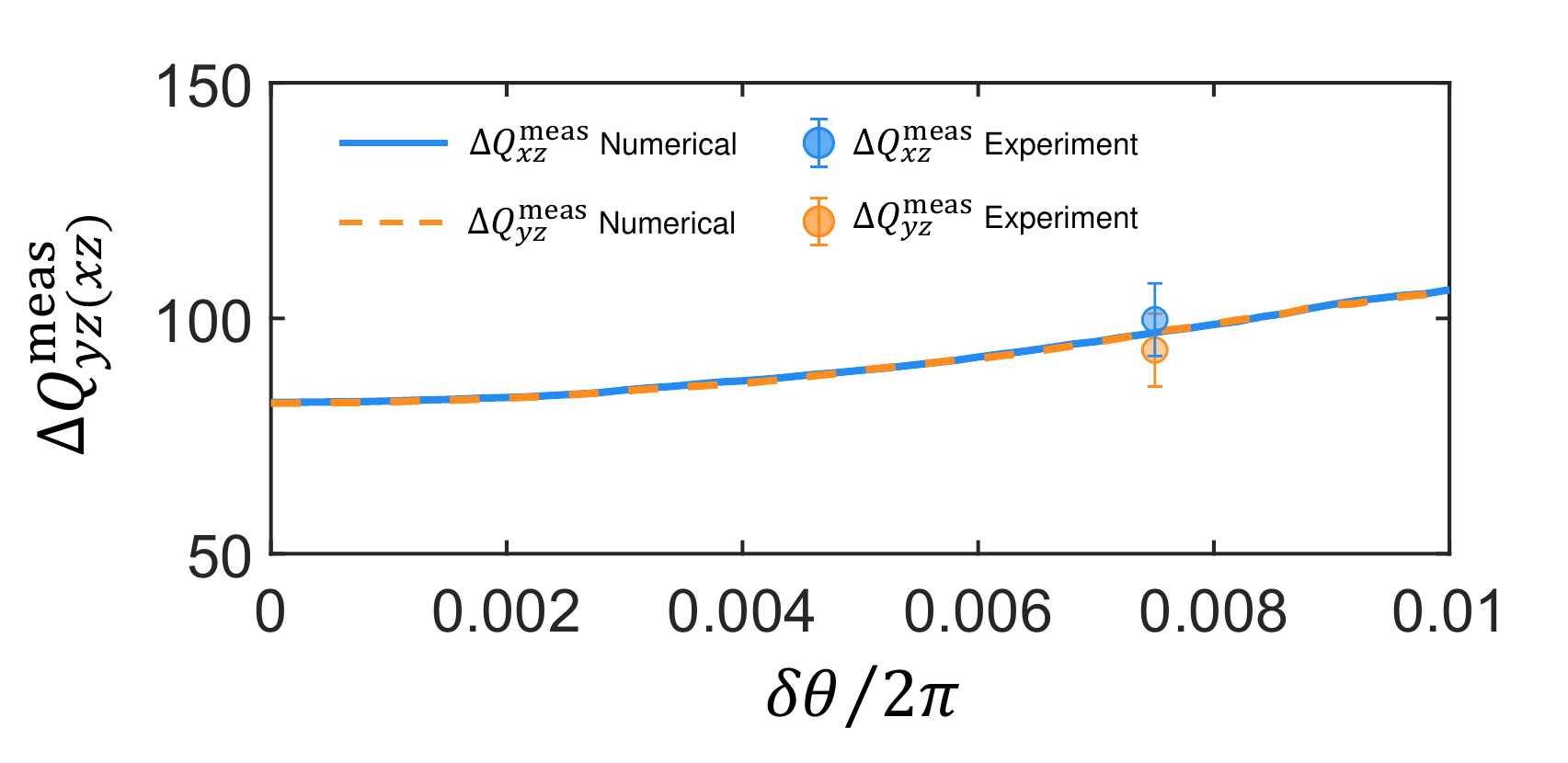}
\caption{Calibration of the pump phase noise strength with spin-nematic squeezed state. The blue solid line represents the numerical simulation of $\Delta Q_{xz}^{\mathrm{meas}}$ as a function of the pump phase noise strength, while the orange dashed line corresponds to that of $\Delta Q_{yz}^{\mathrm{meas}}$. The experimental results are depicted by the blue and orange dots, which align closely with the corresponding numerical simulations at $\delta \theta \approx 0.0075 \times 2\pi$.}
\label{fig:pump_calib}
\end{figure}

When measuring squeezing in both spin-nematic subspaces, the phases of the two coherent pumps are required to differ by $\pi/2$. Due to the imperfection of MW pulses, fluctuation of the pump phase needs to be considered as well. While the polar state is insensitive to this noise, spin-nematic squeezed state is extremely sensitive to it, which is modeled by adding Gaussian distributed random phase into the pump mode
\begin{equation}\label{SM:pump_phase_noise}
\begin{aligned}
a_{2,-2}^{\mathrm{out}} =\frac{1}{\sqrt{2}}\big(a_{S}+i\varepsilon_{1}e^{i\theta_{1}}a_{1,0}\big), \\
a_{2,-1}^{\mathrm{out}} =\frac{1}{\sqrt{2}}\big(ia_{S}+\varepsilon_{_1}e^{i\theta_{_1}}a_{_{1,0}}\big), \\
a_{2,+2}^{\mathrm{out}} =\frac{1}{\sqrt{2}}\big(a_{A}-\varepsilon_{2}e^{i\theta_{2}}a_{1,0}\big), \\
a_{2,+1}^{\mathrm{out}} =\frac{1}{\sqrt{2}}\big(a_{A}+\varepsilon_{2}e^{i\theta_{2}}a_{1,0}\big).
\end{aligned}
\end{equation}
Here we assume $\delta \theta_1=\delta \theta_2=\delta \theta$ for simplicity. To calibrate the noise strength, we numerically calculate the fluctuations of the simultaneously measured $Q_{yz}$ and $Q_{xz}$ for the spin-nematic squeezed state. As shown in Fig.~\ref{fig:pump_calib}, our numerical simulations corroborate well with experimental results at $\delta \theta \approx 0.0075\times2\pi$.

\section{Analysis of the metrological gain}
\subsection{Phase sensitivities}\label{3A}
We employ the error propagation formula to analyze the phase sensitivity of our joint parameter estimation protocol. The sensitivities of $\varphi_1$ and $\varphi_2$ can be obtained with
\begin{equation}\label{SM:sensitivity}
\begin{aligned}
\Delta\varphi_1&=\frac{\Delta Q_{yz}}{\delta\left\langle Q_{yz}\right\rangle/\delta\varphi_1}=\frac{\Delta  Q_{yz}}{\delta\left\langle Q_{yz}\right\rangle/\left[(\cos\Theta) \delta\alpha+\alpha(\sin\Theta) \delta\Theta\right]},\\
\Delta\varphi_2&=\frac{\Delta Q_{xz}}{\delta\left\langle Q_{xz}\right\rangle/\delta\varphi_2}=\frac{\Delta Q_{xz}}{\delta\left\langle Q_{xz}\right\rangle/\left[(\sin\Theta) \delta\alpha+\alpha(\cos\Theta)\delta\Theta\right]}.
\end{aligned}
\end{equation}
The phases to be estimated $\varphi_1$ and $\varphi_2$ are encoded in the alternatively parameterized form $\varphi_1=\alpha \cos \Theta$ and $\varphi_2=\alpha \sin \Theta$, where $\alpha$ of the imprinted phase is controlled experimentally by the duration of the RF pulse while $\Theta$ by the RF phase.

\subsection{Standard quantum limit of two-phase joint estimation}\label{3B}
The quantum metrological gain for parameter estimation protocol is customarily discussed with respect to its classical reference, the standard quantum limit (SQL), which is defined as the optimal phase sensitivity that can be achieved without quantum entanglement. When the unentangled polar state is employed as the probe, the sensitivity for a single parameter estimation is given by
\begin{equation}
\Delta\varphi_\mathrm{SQL}=\frac{1}{2\sqrt{N}},
\end{equation}
where the factor of 2 arises from the presence of three interference modes (or three paths)~\cite{liu2022nonlinear,mao2023quantum}. 
In the present work, two phases are estimated simultaneously, leading to the SQL as
\begin{equation}\label{SM:SQL}
\Delta\vec{\varphi}_{\mathrm{SQL}}=\sqrt{\left(\Delta\varphi_{1,\mathrm{SQL}}\right)^{2}+\left(\Delta\varphi_{2,\mathrm{SQL}}\right)^{2}}=\frac{1}{\sqrt{2N}}.
\end{equation}
\begin{figure*}[!htp]
  \centering
  \includegraphics[width=1.0\linewidth]{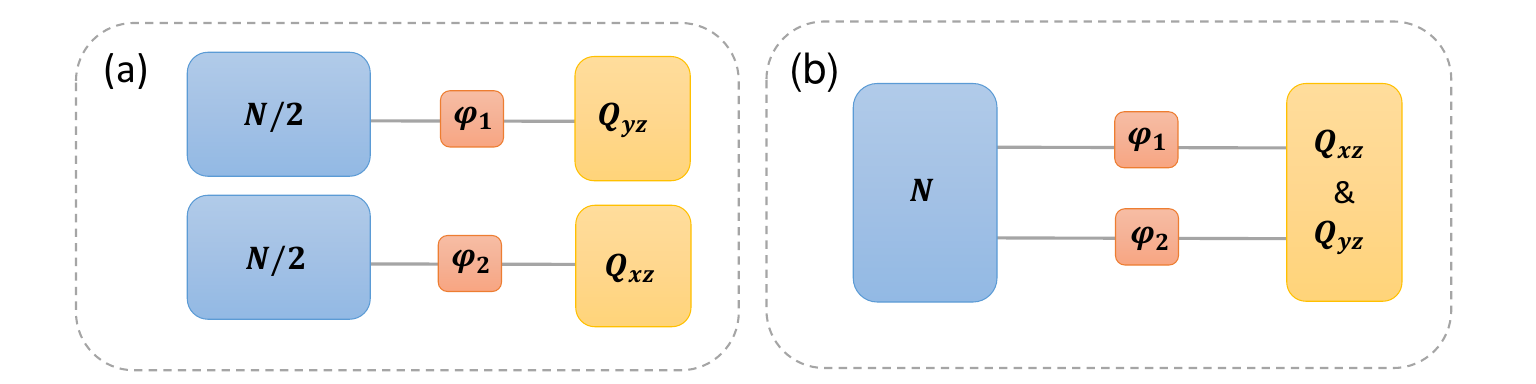}
  \caption{Two distinct scenarios in two-parameter estiamtion. (a) $N$ atoms are divided into two halves with each group estimating a single phase. (b)Every one of the $N$ atoms are encoded with two phases simultaneously.}
  \label{fig:scenario}
\end{figure*}

It should be noted that the SQL Eq.~\eqref{SM:SQL} corresponds to the scenario depicted in the Fig.~\ref{fig:scenario}(b) below, where each atom is encoded with two phases, and is measured simultaneously, i.e., within one shot of
absorption imaging. In this scenario, every atom contributes to the estimation of both parameters. In the main tex, the metrological gain is referenced to Eq.~\eqref{SM:SQL}, as our work corresponds to the scenario depicted in Fig.~\ref{fig:scenario}(b).

In the scenario shown in Fig.~\ref{fig:scenario}(a), $N$ atoms are divided into two halves equally, with each estimating a single phase. Consequently, the SQL for each group is given by $\Delta\varphi_{\mathrm{SQL}}=1/{2\sqrt{N/2}}$. Referenced to SQL of this scenario, the metrological gain reported in the main text will be 3 dB higher.

\section{Experimental methods } 
\subsection{Initial state preparation } 
We prepare an initial state with about 26500 $^{87}$Rb atoms condensed in the ground $|F=1, m_F=0\rangle$ hyperfine state, confined in a crossed dipole trap with trapping frequencies around $2\pi \times(190, 90, 120)$ Hz along three orthogonal directions, by first removing atoms occupying other states with a gradient magnetic field. The bias magnetic field is set at 0.838 G, which offers a quadratic magnetic Zeeman shift of $q_B \approx 2\pi \times 50$ Hz $=13|c_2|$, which is sufficiently large to inhibit spin-mixing. To further purify the initial polar state, two resonant microwave  pulses coupling $ |F=1, m_F=\pm1\rangle$ to $|F=2, m_F=\pm2\rangle$ are applied, followed by a probe light beam  pulse to flush away the remaining $|\pm1\rangle$ atoms. Subsequent spin-mixing evolution is initiated by switching on a microwave dressing field 12.18 MHz blue-detuned from the $ |1, 0\rangle$  to $ |2, 0\rangle$ transition, which quenches $q$ from $13|c_2|$ to near $|c_2|$.


\subsection{MW pulse synthesizing}
During the experiment, a sequence of phase-coherent microwave pulses with controllable $\sigma_{+/-}$-polarization is needed. It is synthesized by combining a microwave signal generator (SMB100A) with output frequency set at 6.6 GHz, and a four-channel AWG (DN2.663-04), output set at 200 MHz, using a Mini-Circuits mixer. The power stabilization is achieved by a power feedback loop. The polarization control is achieved by adjusting the relative phase between the two microwave antennas, which are placed along different spatial directions, hence carrying different polarization components. The signals of the two antennas are generated from two AWG channels, and they remain phase coherent.

\subsection{Calibration of atom number detection }
\begin{figure}[!htp]
  \centering
  \includegraphics[width=0.75\linewidth]{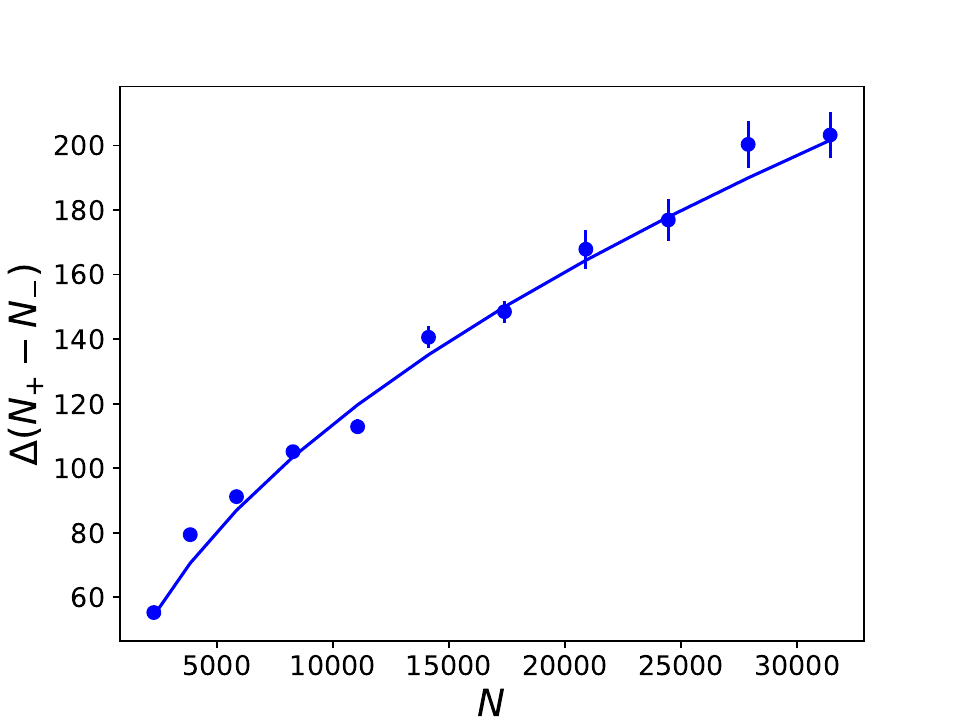}
  \caption{Calibration of the atom number detection accuracy. The solid line denotes $\sqrt{N}$, the theoretical number difference variance of a coherent state. The blue dots show the measured variance, which match well with the solid line, affirming the accuracy of atom number detection.}
  \label{fig:detc}
\end{figure}

The detection signal comes from absorption imaging, therefore the detected atom number accuracy is of great importance to the measurement. We prepare initial polar states with all atoms in $|F=1, m_F=0\rangle$ at different numbers, then apply a resonant RF $\pi/2$ pulse to generate coherent spin state $(|+1\rangle+|-1\rangle)^{\otimes N}/2^{N/2}$ and measure the  variance of number difference between $ |1, \pm1\rangle$ states. As shown in Fig.~{\ref{fig:detc}}, we find the number difference variance matches with that of the coherent state projection noise, which affirms the detection accuracy. 

\subsection{RF amplitude calibration }
\begin{figure}[!htp]
  \centering
  \includegraphics[width=0.75\linewidth]{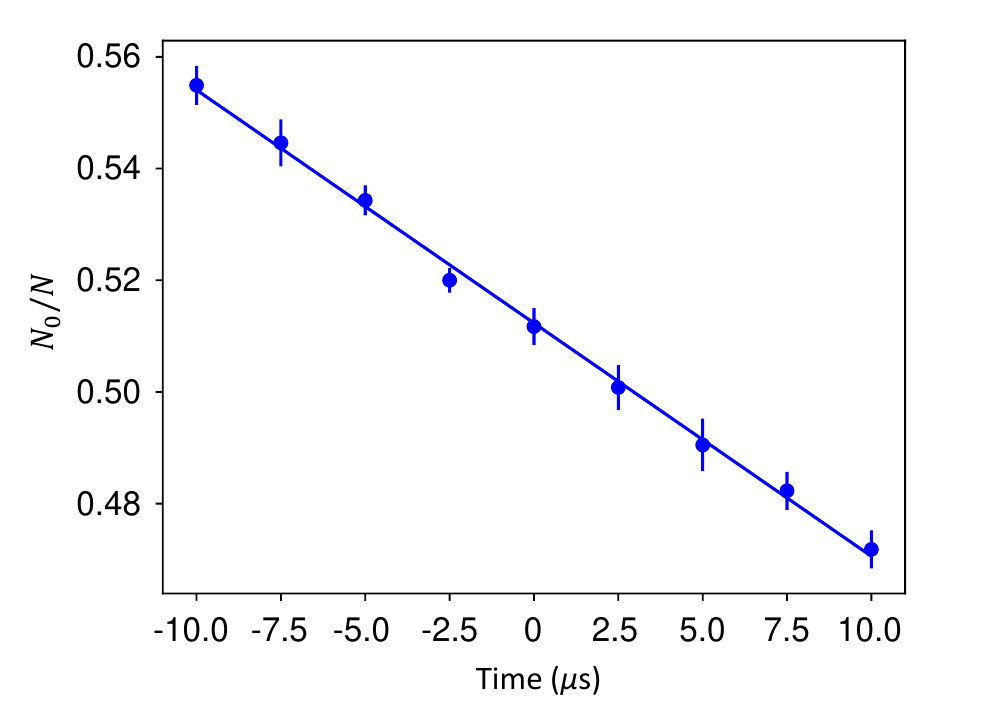}
     \caption{The calibration of RF rotation angle. We use two resonant RF pulses with $\pi$ phase difference to realize small angle rotation. The rotation angle is adjusted by fixing one pulse and changing the duration of the other. The blue dots show the measured $\langle N_0\rangle/N$ changes for coherent state $(-i|+1\rangle+\sqrt{2}|0\rangle-i|-1\rangle)^{\otimes N}/2^{N}$ under such rotations, and the solid line depicts a linear fit of the measured data.}
  \label{fig:calirf}
\end{figure}

The small angle RF rotation is carried out by two resonant RF pulses with $\pi$ phase difference, and the rotation angle is adjusted by changing pulse duration. We prepare the coherent state $(-i|+1\rangle+\sqrt{2}|0\rangle-i|-1\rangle)^{\otimes N}/2^{N}$ with $\langle N_0\rangle/N=1/2$ by RF rotation applied to the initial polar state. Subsequently, a small-angle RF rotation is applied along the same axis, resulting in a change in $\langle N_0\rangle/N$ proportional to the rotation angle $\alpha$, expressed as $\langle N_0\rangle/N=1/2+\alpha$ for small values of $\alpha$. The population changes with different pulse duration, providing a linear calibration to the rotation angle, as shown in Fig.~{\ref{fig:calirf}}.

\begin{thebibliography}{49}%
\makeatletter
\providecommand \@ifxundefined [1]{%
 \@ifx{#1\undefined}
}%
\providecommand \@ifnum [1]{%
 \ifnum #1\expandafter \@firstoftwo
 \else \expandafter \@secondoftwo
 \fi
}%
\providecommand \@ifx [1]{%
 \ifx #1\expandafter \@firstoftwo
 \else \expandafter \@secondoftwo
 \fi
}%
\providecommand \natexlab [1]{#1}%
\providecommand \enquote  [1]{``#1''}%
\providecommand \bibnamefont  [1]{#1}%
\providecommand \bibfnamefont [1]{#1}%
\providecommand \citenamefont [1]{#1}%
\providecommand \href@noop [0]{\@secondoftwo}%
\providecommand \href [0]{\begingroup \@sanitize@url \@href}%
\providecommand \@href[1]{\@@startlink{#1}\@@href}%
\providecommand \@@href[1]{\endgroup#1\@@endlink}%
\providecommand \@sanitize@url [0]{\catcode `\\12\catcode `\$12\catcode
  `\&12\catcode `\#12\catcode `\^12\catcode `\_12\catcode `\%12\relax}%
\providecommand \@@startlink[1]{}%
\providecommand \@@endlink[0]{}%
\providecommand \url  [0]{\begingroup\@sanitize@url \@url }%
\providecommand \@url [1]{\endgroup\@href {#1}{\urlprefix }}%
\providecommand \urlprefix  [0]{URL }%
\providecommand \Eprint [0]{\href }%
\providecommand \doibase [0]{https://doi.org/}%
\providecommand \selectlanguage [0]{\@gobble}%
\providecommand \bibinfo  [0]{\@secondoftwo}%
\providecommand \bibfield  [0]{\@secondoftwo}%
\providecommand \translation [1]{[#1]}%
\providecommand \BibitemOpen [0]{}%
\providecommand \bibitemStop [0]{}%
\providecommand \bibitemNoStop [0]{.\EOS\space}%
\providecommand \EOS [0]{\spacefactor3000\relax}%
\providecommand \BibitemShut  [1]{\csname bibitem#1\endcsname}%
\let\auto@bib@innerbib\@empty
\bibitem [{\citenamefont {Giovannetti}\ \emph {et~al.}(2004)\citenamefont
  {Giovannetti}, \citenamefont {Lloyd},\ and\ \citenamefont
  {Maccone}}]{giovannetti2004quantum}%
  \BibitemOpen
  \bibfield  {author} {\bibinfo {author} {\bibfnamefont {V.}~\bibnamefont
  {Giovannetti}}, \bibinfo {author} {\bibfnamefont {S.}~\bibnamefont {Lloyd}},\
  and\ \bibinfo {author} {\bibfnamefont {L.}~\bibnamefont {Maccone}},\
  }\bibfield  {title} {\bibinfo {title} {Quantum-enhanced measurements: beating
  the standard quantum limit},\ }\href
  {https://www.science.org/doi/10.1126/science.1104149} {\bibfield  {journal}
  {\bibinfo  {journal} {Science}\ }\textbf {\bibinfo {volume} {306}},\ \bibinfo
  {pages} {1330} (\bibinfo {year} {2004})}\BibitemShut {NoStop}%
\bibitem [{\citenamefont {Giovannetti}\ \emph {et~al.}(2011)\citenamefont
  {Giovannetti}, \citenamefont {Lloyd},\ and\ \citenamefont
  {Maccone}}]{giovannetti2011advances}%
  \BibitemOpen
  \bibfield  {author} {\bibinfo {author} {\bibfnamefont {V.}~\bibnamefont
  {Giovannetti}}, \bibinfo {author} {\bibfnamefont {S.}~\bibnamefont {Lloyd}},\
  and\ \bibinfo {author} {\bibfnamefont {L.}~\bibnamefont {Maccone}},\
  }\bibfield  {title} {\bibinfo {title} {Advances in quantum metrology},\
  }\href {https://doi.org/10.1038/nphoton.2011.35} {\bibfield  {journal}
  {\bibinfo  {journal} {Nature photonics}\ }\textbf {\bibinfo {volume} {5}},\
  \bibinfo {pages} {222} (\bibinfo {year} {2011})}\BibitemShut {NoStop}%
\bibitem [{\citenamefont {Pezz\`e}\ \emph {et~al.}(2018)\citenamefont
  {Pezz\`e}, \citenamefont {Smerzi}, \citenamefont {Oberthaler}, \citenamefont
  {Schmied},\ and\ \citenamefont {Treutlein}}]{pezze2018quantum}%
  \BibitemOpen
  \bibfield  {author} {\bibinfo {author} {\bibfnamefont {L.}~\bibnamefont
  {Pezz\`e}}, \bibinfo {author} {\bibfnamefont {A.}~\bibnamefont {Smerzi}},
  \bibinfo {author} {\bibfnamefont {M.~K.}\ \bibnamefont {Oberthaler}},
  \bibinfo {author} {\bibfnamefont {R.}~\bibnamefont {Schmied}},\ and\ \bibinfo
  {author} {\bibfnamefont {P.}~\bibnamefont {Treutlein}},\ }\bibfield  {title}
  {\bibinfo {title} {Quantum metrology with nonclassical states of atomic
  ensembles},\ }\href {https://doi.org/10.1103/RevModPhys.90.035005} {\bibfield
   {journal} {\bibinfo  {journal} {Rev. Mod. Phys.}\ }\textbf {\bibinfo
  {volume} {90}},\ \bibinfo {pages} {035005} (\bibinfo {year}
  {2018})}\BibitemShut {NoStop}%
\bibitem [{\citenamefont {Caves}(1981)}]{caves1981quantum}%
  \BibitemOpen
  \bibfield  {author} {\bibinfo {author} {\bibfnamefont {C.~M.}\ \bibnamefont
  {Caves}},\ }\bibfield  {title} {\bibinfo {title} {Quantum-mechanical noise in
  an interferometer},\ }\href {https://doi.org/10.1103/PhysRevD.23.1693}
  {\bibfield  {journal} {\bibinfo  {journal} {Phys. Rev. D}\ }\textbf {\bibinfo
  {volume} {23}},\ \bibinfo {pages} {1693} (\bibinfo {year}
  {1981})}\BibitemShut {NoStop}%
\bibitem [{\citenamefont {Aasi}\ \emph {et~al.}(2013)\citenamefont {Aasi},
  \citenamefont {Abadie}, \citenamefont {Abbott}, \citenamefont {Abbott},
  \citenamefont {Abbott}, \citenamefont {Abernathy}, \citenamefont {Adams},
  \citenamefont {Adams}, \citenamefont {Addesso}, \citenamefont {Adhikari}
  \emph {et~al.}}]{aasi2013enhanced}%
  \BibitemOpen
  \bibfield  {author} {\bibinfo {author} {\bibfnamefont {J.}~\bibnamefont
  {Aasi}}, \bibinfo {author} {\bibfnamefont {J.}~\bibnamefont {Abadie}},
  \bibinfo {author} {\bibfnamefont {B.}~\bibnamefont {Abbott}}, \bibinfo
  {author} {\bibfnamefont {R.}~\bibnamefont {Abbott}}, \bibinfo {author}
  {\bibfnamefont {T.}~\bibnamefont {Abbott}}, \bibinfo {author} {\bibfnamefont
  {M.}~\bibnamefont {Abernathy}}, \bibinfo {author} {\bibfnamefont
  {C.}~\bibnamefont {Adams}}, \bibinfo {author} {\bibfnamefont
  {T.}~\bibnamefont {Adams}}, \bibinfo {author} {\bibfnamefont
  {P.}~\bibnamefont {Addesso}}, \bibinfo {author} {\bibfnamefont
  {R.}~\bibnamefont {Adhikari}}, \emph {et~al.},\ }\bibfield  {title} {\bibinfo
  {title} {Enhanced sensitivity of the ligo gravitational wave detector by
  using squeezed states of light},\ }\href
  {https://doi.org/10.1038/nphoton.2013.177} {\bibfield  {journal} {\bibinfo
  {journal} {Nature Photonics}\ }\textbf {\bibinfo {volume} {7}},\ \bibinfo
  {pages} {613} (\bibinfo {year} {2013})}\BibitemShut {NoStop}%
\bibitem [{\citenamefont {Polino}\ \emph {et~al.}(2020)\citenamefont {Polino},
  \citenamefont {Valeri}, \citenamefont {Spagnolo},\ and\ \citenamefont
  {Sciarrino}}]{polino2020photonic}%
  \BibitemOpen
  \bibfield  {author} {\bibinfo {author} {\bibfnamefont {E.}~\bibnamefont
  {Polino}}, \bibinfo {author} {\bibfnamefont {M.}~\bibnamefont {Valeri}},
  \bibinfo {author} {\bibfnamefont {N.}~\bibnamefont {Spagnolo}},\ and\
  \bibinfo {author} {\bibfnamefont {F.}~\bibnamefont {Sciarrino}},\ }\bibfield
  {title} {\bibinfo {title} {Photonic quantum metrology},\ }\href
  {https://doi.org/10.1116/5.0007577} {\bibfield  {journal} {\bibinfo
  {journal} {AVS Quantum Science}\ }\textbf {\bibinfo {volume} {2}} (\bibinfo
  {year} {2020})}\BibitemShut {NoStop}%
\bibitem [{\citenamefont {Appel}\ \emph {et~al.}(2009)\citenamefont {Appel},
  \citenamefont {Windpassinger}, \citenamefont {Oblak}, \citenamefont {Hoff},
  \citenamefont {Kj{\ae}rgaard},\ and\ \citenamefont
  {Polzik}}]{appel2009mesoscopic}%
  \BibitemOpen
  \bibfield  {author} {\bibinfo {author} {\bibfnamefont {J.}~\bibnamefont
  {Appel}}, \bibinfo {author} {\bibfnamefont {P.~J.}\ \bibnamefont
  {Windpassinger}}, \bibinfo {author} {\bibfnamefont {D.}~\bibnamefont
  {Oblak}}, \bibinfo {author} {\bibfnamefont {U.~B.}\ \bibnamefont {Hoff}},
  \bibinfo {author} {\bibfnamefont {N.}~\bibnamefont {Kj{\ae}rgaard}},\ and\
  \bibinfo {author} {\bibfnamefont {E.~S.}\ \bibnamefont {Polzik}},\ }\bibfield
   {title} {\bibinfo {title} {Mesoscopic atomic entanglement for precision
  measurements beyond the standard quantum limit},\ }\href
  {https://www.pnas.org/doi/abs/10.1073/pnas.0901550106} {\bibfield  {journal}
  {\bibinfo  {journal} {Proceedings of the National Academy of Sciences}\
  }\textbf {\bibinfo {volume} {106}},\ \bibinfo {pages} {10960} (\bibinfo
  {year} {2009})}\BibitemShut {NoStop}%
\bibitem [{\citenamefont {Schleier-Smith}\ \emph {et~al.}(2010)\citenamefont
  {Schleier-Smith}, \citenamefont {Leroux},\ and\ \citenamefont
  {Vuleti\ifmmode~\acute{c}\else \'{c}\fi{}}}]{schleier2010states}%
  \BibitemOpen
  \bibfield  {author} {\bibinfo {author} {\bibfnamefont {M.~H.}\ \bibnamefont
  {Schleier-Smith}}, \bibinfo {author} {\bibfnamefont {I.~D.}\ \bibnamefont
  {Leroux}},\ and\ \bibinfo {author} {\bibfnamefont {V.}~\bibnamefont
  {Vuleti\ifmmode~\acute{c}\else \'{c}\fi{}}},\ }\bibfield  {title} {\bibinfo
  {title} {States of an ensemble of two-level atoms with reduced quantum
  uncertainty},\ }\href {https://doi.org/10.1103/PhysRevLett.104.073604}
  {\bibfield  {journal} {\bibinfo  {journal} {Phys. Rev. Lett.}\ }\textbf
  {\bibinfo {volume} {104}},\ \bibinfo {pages} {073604} (\bibinfo {year}
  {2010})}\BibitemShut {NoStop}%
\bibitem [{\citenamefont {Gross}\ \emph {et~al.}(2010)\citenamefont {Gross},
  \citenamefont {Zibold}, \citenamefont {Nicklas}, \citenamefont {Esteve},\
  and\ \citenamefont {Oberthaler}}]{gross2010nonlinear}%
  \BibitemOpen
  \bibfield  {author} {\bibinfo {author} {\bibfnamefont {C.}~\bibnamefont
  {Gross}}, \bibinfo {author} {\bibfnamefont {T.}~\bibnamefont {Zibold}},
  \bibinfo {author} {\bibfnamefont {E.}~\bibnamefont {Nicklas}}, \bibinfo
  {author} {\bibfnamefont {J.}~\bibnamefont {Esteve}},\ and\ \bibinfo {author}
  {\bibfnamefont {M.~K.}\ \bibnamefont {Oberthaler}},\ }\bibfield  {title}
  {\bibinfo {title} {Nonlinear atom interferometer surpasses classical
  precision limit},\ }\href {https://doi.org/10.1038/nature08919} {\bibfield
  {journal} {\bibinfo  {journal} {Nature}\ }\textbf {\bibinfo {volume} {464}},\
  \bibinfo {pages} {1165} (\bibinfo {year} {2010})}\BibitemShut {NoStop}%
\bibitem [{\citenamefont {Riedel}\ \emph {et~al.}(2010)\citenamefont {Riedel},
  \citenamefont {B{\"o}hi}, \citenamefont {Li}, \citenamefont {H{\"a}nsch},
  \citenamefont {Sinatra},\ and\ \citenamefont {Treutlein}}]{riedel2010atom}%
  \BibitemOpen
  \bibfield  {author} {\bibinfo {author} {\bibfnamefont {M.~F.}\ \bibnamefont
  {Riedel}}, \bibinfo {author} {\bibfnamefont {P.}~\bibnamefont {B{\"o}hi}},
  \bibinfo {author} {\bibfnamefont {Y.}~\bibnamefont {Li}}, \bibinfo {author}
  {\bibfnamefont {T.~W.}\ \bibnamefont {H{\"a}nsch}}, \bibinfo {author}
  {\bibfnamefont {A.}~\bibnamefont {Sinatra}},\ and\ \bibinfo {author}
  {\bibfnamefont {P.}~\bibnamefont {Treutlein}},\ }\bibfield  {title} {\bibinfo
  {title} {Atom-chip-based generation of entanglement for quantum metrology},\
  }\href {https://doi.org/10.1038/nature08988} {\bibfield  {journal} {\bibinfo
  {journal} {Nature}\ }\textbf {\bibinfo {volume} {464}},\ \bibinfo {pages}
  {1170} (\bibinfo {year} {2010})}\BibitemShut {NoStop}%
\bibitem [{\citenamefont {L{\"u}cke}\ \emph {et~al.}(2011)\citenamefont
  {L{\"u}cke}, \citenamefont {Scherer}, \citenamefont {Kruse}, \citenamefont
  {Pezz{\'e}}, \citenamefont {Deuretzbacher}, \citenamefont {Hyllus},
  \citenamefont {Topic}, \citenamefont {Peise}, \citenamefont {Ertmer},
  \citenamefont {Arlt} \emph {et~al.}}]{lucke2011twin}%
  \BibitemOpen
  \bibfield  {author} {\bibinfo {author} {\bibfnamefont {B.}~\bibnamefont
  {L{\"u}cke}}, \bibinfo {author} {\bibfnamefont {M.}~\bibnamefont {Scherer}},
  \bibinfo {author} {\bibfnamefont {J.}~\bibnamefont {Kruse}}, \bibinfo
  {author} {\bibfnamefont {L.}~\bibnamefont {Pezz{\'e}}}, \bibinfo {author}
  {\bibfnamefont {F.}~\bibnamefont {Deuretzbacher}}, \bibinfo {author}
  {\bibfnamefont {P.}~\bibnamefont {Hyllus}}, \bibinfo {author} {\bibfnamefont
  {O.}~\bibnamefont {Topic}}, \bibinfo {author} {\bibfnamefont
  {J.}~\bibnamefont {Peise}}, \bibinfo {author} {\bibfnamefont
  {W.}~\bibnamefont {Ertmer}}, \bibinfo {author} {\bibfnamefont
  {J.}~\bibnamefont {Arlt}}, \emph {et~al.},\ }\bibfield  {title} {\bibinfo
  {title} {Twin matter waves for interferometry beyond the classical limit},\
  }\href {https://www.science.org/doi/abs/10.1126/science.1208798} {\bibfield
  {journal} {\bibinfo  {journal} {Science}\ }\textbf {\bibinfo {volume}
  {334}},\ \bibinfo {pages} {773} (\bibinfo {year} {2011})}\BibitemShut
  {NoStop}%
\bibitem [{\citenamefont {Sewell}\ \emph {et~al.}(2012)\citenamefont {Sewell},
  \citenamefont {Koschorreck}, \citenamefont {Napolitano}, \citenamefont
  {Dubost}, \citenamefont {Behbood},\ and\ \citenamefont
  {Mitchell}}]{sewell2012magnetic}%
  \BibitemOpen
  \bibfield  {author} {\bibinfo {author} {\bibfnamefont {R.~J.}\ \bibnamefont
  {Sewell}}, \bibinfo {author} {\bibfnamefont {M.}~\bibnamefont {Koschorreck}},
  \bibinfo {author} {\bibfnamefont {M.}~\bibnamefont {Napolitano}}, \bibinfo
  {author} {\bibfnamefont {B.}~\bibnamefont {Dubost}}, \bibinfo {author}
  {\bibfnamefont {N.}~\bibnamefont {Behbood}},\ and\ \bibinfo {author}
  {\bibfnamefont {M.~W.}\ \bibnamefont {Mitchell}},\ }\bibfield  {title}
  {\bibinfo {title} {Magnetic sensitivity beyond the projection noise limit by
  spin squeezing},\ }\href {https://doi.org/10.1103/PhysRevLett.109.253605}
  {\bibfield  {journal} {\bibinfo  {journal} {Phys. Rev. Lett.}\ }\textbf
  {\bibinfo {volume} {109}},\ \bibinfo {pages} {253605} (\bibinfo {year}
  {2012})}\BibitemShut {NoStop}%
\bibitem [{\citenamefont {Bohnet}\ \emph {et~al.}(2014)\citenamefont {Bohnet},
  \citenamefont {Cox}, \citenamefont {Norcia}, \citenamefont {Weiner},
  \citenamefont {Chen},\ and\ \citenamefont {Thompson}}]{bohnet2014reduced}%
  \BibitemOpen
  \bibfield  {author} {\bibinfo {author} {\bibfnamefont {J.~G.}\ \bibnamefont
  {Bohnet}}, \bibinfo {author} {\bibfnamefont {K.~C.}\ \bibnamefont {Cox}},
  \bibinfo {author} {\bibfnamefont {M.~A.}\ \bibnamefont {Norcia}}, \bibinfo
  {author} {\bibfnamefont {J.~M.}\ \bibnamefont {Weiner}}, \bibinfo {author}
  {\bibfnamefont {Z.}~\bibnamefont {Chen}},\ and\ \bibinfo {author}
  {\bibfnamefont {J.~K.}\ \bibnamefont {Thompson}},\ }\bibfield  {title}
  {\bibinfo {title} {Reduced spin measurement back-action for a phase
  sensitivity ten times beyond the standard quantum limit},\ }\href
  {https://doi.org/10.1038/nphoton.2014.151} {\bibfield  {journal} {\bibinfo
  {journal} {Nature Photonics}\ }\textbf {\bibinfo {volume} {8}},\ \bibinfo
  {pages} {731} (\bibinfo {year} {2014})}\BibitemShut {NoStop}%
\bibitem [{\citenamefont {Hosten}\ \emph {et~al.}(2016)\citenamefont {Hosten},
  \citenamefont {Engelsen}, \citenamefont {Krishnakumar},\ and\ \citenamefont
  {Kasevich}}]{hosten2016measurement}%
  \BibitemOpen
  \bibfield  {author} {\bibinfo {author} {\bibfnamefont {O.}~\bibnamefont
  {Hosten}}, \bibinfo {author} {\bibfnamefont {N.~J.}\ \bibnamefont
  {Engelsen}}, \bibinfo {author} {\bibfnamefont {R.}~\bibnamefont
  {Krishnakumar}},\ and\ \bibinfo {author} {\bibfnamefont {M.~A.}\ \bibnamefont
  {Kasevich}},\ }\bibfield  {title} {\bibinfo {title} {Measurement noise 100
  times lower than the quantum-projection limit using entangled atoms},\ }\href
  {https://doi.org/10.1038/nature16176} {\bibfield  {journal} {\bibinfo
  {journal} {Nature}\ }\textbf {\bibinfo {volume} {529}},\ \bibinfo {pages}
  {505} (\bibinfo {year} {2016})}\BibitemShut {NoStop}%
\bibitem [{\citenamefont {Luo}\ \emph {et~al.}(2017)\citenamefont {Luo},
  \citenamefont {Zou}, \citenamefont {Wu}, \citenamefont {Liu}, \citenamefont
  {Han}, \citenamefont {Tey},\ and\ \citenamefont
  {You}}]{luo2017deterministic}%
  \BibitemOpen
  \bibfield  {author} {\bibinfo {author} {\bibfnamefont {X.-Y.}\ \bibnamefont
  {Luo}}, \bibinfo {author} {\bibfnamefont {Y.-Q.}\ \bibnamefont {Zou}},
  \bibinfo {author} {\bibfnamefont {L.-N.}\ \bibnamefont {Wu}}, \bibinfo
  {author} {\bibfnamefont {Q.}~\bibnamefont {Liu}}, \bibinfo {author}
  {\bibfnamefont {M.-F.}\ \bibnamefont {Han}}, \bibinfo {author} {\bibfnamefont
  {M.~K.}\ \bibnamefont {Tey}},\ and\ \bibinfo {author} {\bibfnamefont
  {L.}~\bibnamefont {You}},\ }\bibfield  {title} {\bibinfo {title}
  {Deterministic entanglement generation from driving through quantum phase
  transitions},\ }\href {https://www.science.org/doi/10.1126/science.aag1106}
  {\bibfield  {journal} {\bibinfo  {journal} {Science}\ }\textbf {\bibinfo
  {volume} {355}},\ \bibinfo {pages} {620} (\bibinfo {year}
  {2017})}\BibitemShut {NoStop}%
\bibitem [{\citenamefont {Colombo}\ \emph {et~al.}(2022)\citenamefont
  {Colombo}, \citenamefont {Pedrozo-Pe{\~n}afiel}, \citenamefont {Adiyatullin},
  \citenamefont {Li}, \citenamefont {Mendez}, \citenamefont {Shu},\ and\
  \citenamefont {Vuleti{\'c}}}]{colombo2022time}%
  \BibitemOpen
  \bibfield  {author} {\bibinfo {author} {\bibfnamefont {S.}~\bibnamefont
  {Colombo}}, \bibinfo {author} {\bibfnamefont {E.}~\bibnamefont
  {Pedrozo-Pe{\~n}afiel}}, \bibinfo {author} {\bibfnamefont {A.~F.}\
  \bibnamefont {Adiyatullin}}, \bibinfo {author} {\bibfnamefont
  {Z.}~\bibnamefont {Li}}, \bibinfo {author} {\bibfnamefont {E.}~\bibnamefont
  {Mendez}}, \bibinfo {author} {\bibfnamefont {C.}~\bibnamefont {Shu}},\ and\
  \bibinfo {author} {\bibfnamefont {V.}~\bibnamefont {Vuleti{\'c}}},\
  }\bibfield  {title} {\bibinfo {title} {Time-reversal-based quantum metrology
  with many-body entangled states},\ }\href
  {https://doi.org/10.1038/s41567-022-01653-5} {\bibfield  {journal} {\bibinfo
  {journal} {Nature Physics}\ }\textbf {\bibinfo {volume} {18}},\ \bibinfo
  {pages} {925} (\bibinfo {year} {2022})}\BibitemShut {NoStop}%
\bibitem [{\citenamefont {Gilmore}\ \emph {et~al.}(2021)\citenamefont
  {Gilmore}, \citenamefont {Affolter}, \citenamefont {Lewis-Swan},
  \citenamefont {Barberena}, \citenamefont {Jordan}, \citenamefont {Rey},\ and\
  \citenamefont {Bollinger}}]{gilmore2021quantum}%
  \BibitemOpen
  \bibfield  {author} {\bibinfo {author} {\bibfnamefont {K.~A.}\ \bibnamefont
  {Gilmore}}, \bibinfo {author} {\bibfnamefont {M.}~\bibnamefont {Affolter}},
  \bibinfo {author} {\bibfnamefont {R.~J.}\ \bibnamefont {Lewis-Swan}},
  \bibinfo {author} {\bibfnamefont {D.}~\bibnamefont {Barberena}}, \bibinfo
  {author} {\bibfnamefont {E.}~\bibnamefont {Jordan}}, \bibinfo {author}
  {\bibfnamefont {A.~M.}\ \bibnamefont {Rey}},\ and\ \bibinfo {author}
  {\bibfnamefont {J.~J.}\ \bibnamefont {Bollinger}},\ }\bibfield  {title}
  {\bibinfo {title} {Quantum-enhanced sensing of displacements and electric
  fields with two-dimensional trapped-ion crystals},\ }\href
  {https://www.science.org/doi/10.1126/science.abi5226} {\bibfield  {journal}
  {\bibinfo  {journal} {Science}\ }\textbf {\bibinfo {volume} {373}},\ \bibinfo
  {pages} {673} (\bibinfo {year} {2021})}\BibitemShut {NoStop}%
\bibitem [{\citenamefont {Marciniak}\ \emph {et~al.}(2022)\citenamefont
  {Marciniak}, \citenamefont {Feldker}, \citenamefont {Pogorelov},
  \citenamefont {Kaubruegger}, \citenamefont {Vasilyev}, \citenamefont {van
  Bijnen}, \citenamefont {Schindler}, \citenamefont {Zoller}, \citenamefont
  {Blatt},\ and\ \citenamefont {Monz}}]{marciniak2022optimal}%
  \BibitemOpen
  \bibfield  {author} {\bibinfo {author} {\bibfnamefont {C.~D.}\ \bibnamefont
  {Marciniak}}, \bibinfo {author} {\bibfnamefont {T.}~\bibnamefont {Feldker}},
  \bibinfo {author} {\bibfnamefont {I.}~\bibnamefont {Pogorelov}}, \bibinfo
  {author} {\bibfnamefont {R.}~\bibnamefont {Kaubruegger}}, \bibinfo {author}
  {\bibfnamefont {D.~V.}\ \bibnamefont {Vasilyev}}, \bibinfo {author}
  {\bibfnamefont {R.}~\bibnamefont {van Bijnen}}, \bibinfo {author}
  {\bibfnamefont {P.}~\bibnamefont {Schindler}}, \bibinfo {author}
  {\bibfnamefont {P.}~\bibnamefont {Zoller}}, \bibinfo {author} {\bibfnamefont
  {R.}~\bibnamefont {Blatt}},\ and\ \bibinfo {author} {\bibfnamefont
  {T.}~\bibnamefont {Monz}},\ }\bibfield  {title} {\bibinfo {title} {Optimal
  metrology with programmable quantum sensors},\ }\href
  {https://doi.org/10.1038/s41586-022-04435-4} {\bibfield  {journal} {\bibinfo
  {journal} {Nature}\ }\textbf {\bibinfo {volume} {603}},\ \bibinfo {pages}
  {604} (\bibinfo {year} {2022})}\BibitemShut {NoStop}%
\bibitem [{\citenamefont {Wang}\ \emph {et~al.}(2019)\citenamefont {Wang},
  \citenamefont {Wu}, \citenamefont {Ma}, \citenamefont {Cai}, \citenamefont
  {Hu}, \citenamefont {Mu}, \citenamefont {Xu}, \citenamefont {Chen},
  \citenamefont {Wang}, \citenamefont {Song} \emph
  {et~al.}}]{wang2019heisenberg}%
  \BibitemOpen
  \bibfield  {author} {\bibinfo {author} {\bibfnamefont {W.}~\bibnamefont
  {Wang}}, \bibinfo {author} {\bibfnamefont {Y.}~\bibnamefont {Wu}}, \bibinfo
  {author} {\bibfnamefont {Y.}~\bibnamefont {Ma}}, \bibinfo {author}
  {\bibfnamefont {W.}~\bibnamefont {Cai}}, \bibinfo {author} {\bibfnamefont
  {L.}~\bibnamefont {Hu}}, \bibinfo {author} {\bibfnamefont {X.}~\bibnamefont
  {Mu}}, \bibinfo {author} {\bibfnamefont {Y.}~\bibnamefont {Xu}}, \bibinfo
  {author} {\bibfnamefont {Z.-J.}\ \bibnamefont {Chen}}, \bibinfo {author}
  {\bibfnamefont {H.}~\bibnamefont {Wang}}, \bibinfo {author} {\bibfnamefont
  {Y.}~\bibnamefont {Song}}, \emph {et~al.},\ }\bibfield  {title} {\bibinfo
  {title} {Heisenberg-limited single-mode quantum metrology in a
  superconducting circuit},\ }\href
  {https://doi.org/10.1038/s41467-019-12290-7} {\bibfield  {journal} {\bibinfo
  {journal} {Nature Communications}\ }\textbf {\bibinfo {volume} {10}},\
  \bibinfo {pages} {4382} (\bibinfo {year} {2019})}\BibitemShut {NoStop}%
\bibitem [{\citenamefont {Pedrozo-Pe{\~n}afiel}\ \emph
  {et~al.}(2020)\citenamefont {Pedrozo-Pe{\~n}afiel}, \citenamefont {Colombo},
  \citenamefont {Shu}, \citenamefont {Adiyatullin}, \citenamefont {Li},
  \citenamefont {Mendez}, \citenamefont {Braverman}, \citenamefont {Kawasaki},
  \citenamefont {Akamatsu}, \citenamefont {Xiao} \emph
  {et~al.}}]{pedrozo2020entanglement}%
  \BibitemOpen
  \bibfield  {author} {\bibinfo {author} {\bibfnamefont {E.}~\bibnamefont
  {Pedrozo-Pe{\~n}afiel}}, \bibinfo {author} {\bibfnamefont {S.}~\bibnamefont
  {Colombo}}, \bibinfo {author} {\bibfnamefont {C.}~\bibnamefont {Shu}},
  \bibinfo {author} {\bibfnamefont {A.~F.}\ \bibnamefont {Adiyatullin}},
  \bibinfo {author} {\bibfnamefont {Z.}~\bibnamefont {Li}}, \bibinfo {author}
  {\bibfnamefont {E.}~\bibnamefont {Mendez}}, \bibinfo {author} {\bibfnamefont
  {B.}~\bibnamefont {Braverman}}, \bibinfo {author} {\bibfnamefont
  {A.}~\bibnamefont {Kawasaki}}, \bibinfo {author} {\bibfnamefont
  {D.}~\bibnamefont {Akamatsu}}, \bibinfo {author} {\bibfnamefont
  {Y.}~\bibnamefont {Xiao}}, \emph {et~al.},\ }\bibfield  {title} {\bibinfo
  {title} {Entanglement on an optical atomic-clock transition},\ }\href
  {https://doi.org/10.1038/s41586-020-3006-1} {\bibfield  {journal} {\bibinfo
  {journal} {Nature}\ }\textbf {\bibinfo {volume} {588}},\ \bibinfo {pages}
  {414} (\bibinfo {year} {2020})}\BibitemShut {NoStop}%
\bibitem [{\citenamefont {Muessel}\ \emph {et~al.}(2014)\citenamefont
  {Muessel}, \citenamefont {Strobel}, \citenamefont {Linnemann}, \citenamefont
  {Hume},\ and\ \citenamefont {Oberthaler}}]{muessel2014scalable}%
  \BibitemOpen
  \bibfield  {author} {\bibinfo {author} {\bibfnamefont {W.}~\bibnamefont
  {Muessel}}, \bibinfo {author} {\bibfnamefont {H.}~\bibnamefont {Strobel}},
  \bibinfo {author} {\bibfnamefont {D.}~\bibnamefont {Linnemann}}, \bibinfo
  {author} {\bibfnamefont {D.~B.}\ \bibnamefont {Hume}},\ and\ \bibinfo
  {author} {\bibfnamefont {M.~K.}\ \bibnamefont {Oberthaler}},\ }\bibfield
  {title} {\bibinfo {title} {Scalable spin squeezing for quantum-enhanced
  magnetometry with bose-einstein condensates},\ }\href
  {https://doi.org/10.1103/PhysRevLett.113.103004} {\bibfield  {journal}
  {\bibinfo  {journal} {Phys. Rev. Lett.}\ }\textbf {\bibinfo {volume} {113}},\
  \bibinfo {pages} {103004} (\bibinfo {year} {2014})}\BibitemShut {NoStop}%
\bibitem [{\citenamefont {Szczykulska}\ \emph {et~al.}(2016)\citenamefont
  {Szczykulska}, \citenamefont {Baumgratz},\ and\ \citenamefont
  {Datta}}]{szczykulska2016multi}%
  \BibitemOpen
  \bibfield  {author} {\bibinfo {author} {\bibfnamefont {M.}~\bibnamefont
  {Szczykulska}}, \bibinfo {author} {\bibfnamefont {T.}~\bibnamefont
  {Baumgratz}},\ and\ \bibinfo {author} {\bibfnamefont {A.}~\bibnamefont
  {Datta}},\ }\bibfield  {title} {\bibinfo {title} {Multi-parameter quantum
  metrology},\ }\href {https://doi.org/10.1080/23746149.2016.1230476}
  {\bibfield  {journal} {\bibinfo  {journal} {Advances in Physics: X}\ }\textbf
  {\bibinfo {volume} {1}},\ \bibinfo {pages} {621} (\bibinfo {year}
  {2016})}\BibitemShut {NoStop}%
\bibitem [{\citenamefont {Liu}\ \emph {et~al.}(2020)\citenamefont {Liu},
  \citenamefont {Yuan}, \citenamefont {Lu},\ and\ \citenamefont
  {Wang}}]{liu2020quantum}%
  \BibitemOpen
  \bibfield  {author} {\bibinfo {author} {\bibfnamefont {J.}~\bibnamefont
  {Liu}}, \bibinfo {author} {\bibfnamefont {H.}~\bibnamefont {Yuan}}, \bibinfo
  {author} {\bibfnamefont {X.-M.}\ \bibnamefont {Lu}},\ and\ \bibinfo {author}
  {\bibfnamefont {X.}~\bibnamefont {Wang}},\ }\bibfield  {title} {\bibinfo
  {title} {Quantum fisher information matrix and multiparameter estimation},\
  }\href {https://iopscience.iop.org/article/10.1088/1751-8121/ab5d4d}
  {\bibfield  {journal} {\bibinfo  {journal} {Journal of Physics A:
  Mathematical and Theoretical}\ }\textbf {\bibinfo {volume} {53}},\ \bibinfo
  {pages} {023001} (\bibinfo {year} {2020})}\BibitemShut {NoStop}%
\bibitem [{\citenamefont {Albarelli}\ \emph {et~al.}(2020)\citenamefont
  {Albarelli}, \citenamefont {Barbieri}, \citenamefont {Genoni},\ and\
  \citenamefont {Gianani}}]{albarelli2020perspective}%
  \BibitemOpen
  \bibfield  {author} {\bibinfo {author} {\bibfnamefont {F.}~\bibnamefont
  {Albarelli}}, \bibinfo {author} {\bibfnamefont {M.}~\bibnamefont {Barbieri}},
  \bibinfo {author} {\bibfnamefont {M.~G.}\ \bibnamefont {Genoni}},\ and\
  \bibinfo {author} {\bibfnamefont {I.}~\bibnamefont {Gianani}},\ }\bibfield
  {title} {\bibinfo {title} {A perspective on multiparameter quantum metrology:
  From theoretical tools to applications in quantum imaging},\ }\href
  {https://doi.org/10.1016/j.physleta.2020.126311} {\bibfield  {journal}
  {\bibinfo  {journal} {Physics Letters A}\ }\textbf {\bibinfo {volume}
  {384}},\ \bibinfo {pages} {126311} (\bibinfo {year} {2020})}\BibitemShut
  {NoStop}%
\bibitem [{\citenamefont {Bisketzi}\ \emph {et~al.}(2019)\citenamefont
  {Bisketzi}, \citenamefont {Branford},\ and\ \citenamefont
  {Datta}}]{bisketzi2019quantum}%
  \BibitemOpen
  \bibfield  {author} {\bibinfo {author} {\bibfnamefont {E.}~\bibnamefont
  {Bisketzi}}, \bibinfo {author} {\bibfnamefont {D.}~\bibnamefont {Branford}},\
  and\ \bibinfo {author} {\bibfnamefont {A.}~\bibnamefont {Datta}},\ }\bibfield
   {title} {\bibinfo {title} {Quantum limits of localisation microscopy},\
  }\href {https://iopscience.iop.org/article/10.1088/1367-2630/ab58a0}
  {\bibfield  {journal} {\bibinfo  {journal} {New Journal of Physics}\ }\textbf
  {\bibinfo {volume} {21}},\ \bibinfo {pages} {123032} (\bibinfo {year}
  {2019})}\BibitemShut {NoStop}%
\bibitem [{\citenamefont {Baumgratz}\ and\ \citenamefont
  {Datta}(2016)}]{baumgratz2016quantum}%
  \BibitemOpen
  \bibfield  {author} {\bibinfo {author} {\bibfnamefont {T.}~\bibnamefont
  {Baumgratz}}\ and\ \bibinfo {author} {\bibfnamefont {A.}~\bibnamefont
  {Datta}},\ }\bibfield  {title} {\bibinfo {title} {Quantum enhanced estimation
  of a multidimensional field},\ }\href
  {https://doi.org/10.1103/PhysRevLett.116.030801} {\bibfield  {journal}
  {\bibinfo  {journal} {Phys. Rev. Lett.}\ }\textbf {\bibinfo {volume} {116}},\
  \bibinfo {pages} {030801} (\bibinfo {year} {2016})}\BibitemShut {NoStop}%
\bibitem [{\citenamefont {Meng}\ \emph {et~al.}(2023)\citenamefont {Meng},
  \citenamefont {Zhang}, \citenamefont {Zhang}, \citenamefont {Jin},
  \citenamefont {Wang}, \citenamefont {Jiang}, \citenamefont {Xiao},
  \citenamefont {Jia},\ and\ \citenamefont {Xiao}}]{meng2023machine}%
  \BibitemOpen
  \bibfield  {author} {\bibinfo {author} {\bibfnamefont {X.}~\bibnamefont
  {Meng}}, \bibinfo {author} {\bibfnamefont {Y.}~\bibnamefont {Zhang}},
  \bibinfo {author} {\bibfnamefont {X.}~\bibnamefont {Zhang}}, \bibinfo
  {author} {\bibfnamefont {S.}~\bibnamefont {Jin}}, \bibinfo {author}
  {\bibfnamefont {T.}~\bibnamefont {Wang}}, \bibinfo {author} {\bibfnamefont
  {L.}~\bibnamefont {Jiang}}, \bibinfo {author} {\bibfnamefont
  {L.}~\bibnamefont {Xiao}}, \bibinfo {author} {\bibfnamefont {S.}~\bibnamefont
  {Jia}},\ and\ \bibinfo {author} {\bibfnamefont {Y.}~\bibnamefont {Xiao}},\
  }\bibfield  {title} {\bibinfo {title} {Machine learning assisted vector
  atomic magnetometry},\ }\href {https://doi.org/10.1038/s41467-023-41676-x}
  {\bibfield  {journal} {\bibinfo  {journal} {Nature Communications}\ }\textbf
  {\bibinfo {volume} {14}},\ \bibinfo {pages} {6105} (\bibinfo {year}
  {2023})}\BibitemShut {NoStop}%
\bibitem [{\citenamefont {Komar}\ \emph {et~al.}(2014)\citenamefont {Komar},
  \citenamefont {Kessler}, \citenamefont {Bishof}, \citenamefont {Jiang},
  \citenamefont {S{\o}rensen}, \citenamefont {Ye},\ and\ \citenamefont
  {Lukin}}]{komar2014quantum}%
  \BibitemOpen
  \bibfield  {author} {\bibinfo {author} {\bibfnamefont {P.}~\bibnamefont
  {Komar}}, \bibinfo {author} {\bibfnamefont {E.~M.}\ \bibnamefont {Kessler}},
  \bibinfo {author} {\bibfnamefont {M.}~\bibnamefont {Bishof}}, \bibinfo
  {author} {\bibfnamefont {L.}~\bibnamefont {Jiang}}, \bibinfo {author}
  {\bibfnamefont {A.~S.}\ \bibnamefont {S{\o}rensen}}, \bibinfo {author}
  {\bibfnamefont {J.}~\bibnamefont {Ye}},\ and\ \bibinfo {author}
  {\bibfnamefont {M.~D.}\ \bibnamefont {Lukin}},\ }\bibfield  {title} {\bibinfo
  {title} {A quantum network of clocks},\ }\href
  {https://doi.org/10.1038/nphys3000} {\bibfield  {journal} {\bibinfo
  {journal} {Nature Physics}\ }\textbf {\bibinfo {volume} {10}},\ \bibinfo
  {pages} {582} (\bibinfo {year} {2014})}\BibitemShut {NoStop}%
\bibitem [{\citenamefont {Pezz\`e}\ \emph {et~al.}(2017)\citenamefont
  {Pezz\`e}, \citenamefont {Ciampini}, \citenamefont {Spagnolo}, \citenamefont
  {Humphreys}, \citenamefont {Datta}, \citenamefont {Walmsley}, \citenamefont
  {Barbieri}, \citenamefont {Sciarrino},\ and\ \citenamefont
  {Smerzi}}]{pezze2017optimal}%
  \BibitemOpen
  \bibfield  {author} {\bibinfo {author} {\bibfnamefont {L.}~\bibnamefont
  {Pezz\`e}}, \bibinfo {author} {\bibfnamefont {M.~A.}\ \bibnamefont
  {Ciampini}}, \bibinfo {author} {\bibfnamefont {N.}~\bibnamefont {Spagnolo}},
  \bibinfo {author} {\bibfnamefont {P.~C.}\ \bibnamefont {Humphreys}}, \bibinfo
  {author} {\bibfnamefont {A.}~\bibnamefont {Datta}}, \bibinfo {author}
  {\bibfnamefont {I.~A.}\ \bibnamefont {Walmsley}}, \bibinfo {author}
  {\bibfnamefont {M.}~\bibnamefont {Barbieri}}, \bibinfo {author}
  {\bibfnamefont {F.}~\bibnamefont {Sciarrino}},\ and\ \bibinfo {author}
  {\bibfnamefont {A.}~\bibnamefont {Smerzi}},\ }\bibfield  {title} {\bibinfo
  {title} {Optimal measurements for simultaneous quantum estimation of multiple
  phases},\ }\href {https://doi.org/10.1103/PhysRevLett.119.130504} {\bibfield
  {journal} {\bibinfo  {journal} {Phys. Rev. Lett.}\ }\textbf {\bibinfo
  {volume} {119}},\ \bibinfo {pages} {130504} (\bibinfo {year}
  {2017})}\BibitemShut {NoStop}%
\bibitem [{\citenamefont {Conlon}\ \emph {et~al.}(2023)\citenamefont {Conlon},
  \citenamefont {Vogl}, \citenamefont {Marciniak}, \citenamefont {Pogorelov},
  \citenamefont {Yung}, \citenamefont {Eilenberger}, \citenamefont {Berry},
  \citenamefont {Santana}, \citenamefont {Blatt}, \citenamefont {Monz} \emph
  {et~al.}}]{conlon2023approaching}%
  \BibitemOpen
  \bibfield  {author} {\bibinfo {author} {\bibfnamefont {L.~O.}\ \bibnamefont
  {Conlon}}, \bibinfo {author} {\bibfnamefont {T.}~\bibnamefont {Vogl}},
  \bibinfo {author} {\bibfnamefont {C.~D.}\ \bibnamefont {Marciniak}}, \bibinfo
  {author} {\bibfnamefont {I.}~\bibnamefont {Pogorelov}}, \bibinfo {author}
  {\bibfnamefont {S.~K.}\ \bibnamefont {Yung}}, \bibinfo {author}
  {\bibfnamefont {F.}~\bibnamefont {Eilenberger}}, \bibinfo {author}
  {\bibfnamefont {D.~W.}\ \bibnamefont {Berry}}, \bibinfo {author}
  {\bibfnamefont {F.~S.}\ \bibnamefont {Santana}}, \bibinfo {author}
  {\bibfnamefont {R.}~\bibnamefont {Blatt}}, \bibinfo {author} {\bibfnamefont
  {T.}~\bibnamefont {Monz}}, \emph {et~al.},\ }\bibfield  {title} {\bibinfo
  {title} {Approaching optimal entangling collective measurements on quantum
  computing platforms},\ }\href {https://doi.org/10.1038/s41567-022-01875-7}
  {\bibfield  {journal} {\bibinfo  {journal} {Nature Physics}\ }\textbf
  {\bibinfo {volume} {19}},\ \bibinfo {pages} {351} (\bibinfo {year}
  {2023})}\BibitemShut {NoStop}%
\bibitem [{\citenamefont {Braunstein}\ and\ \citenamefont
  {Kimble}(2000)}]{braunstein2000dense}%
  \BibitemOpen
  \bibfield  {author} {\bibinfo {author} {\bibfnamefont {S.~L.}\ \bibnamefont
  {Braunstein}}\ and\ \bibinfo {author} {\bibfnamefont {H.~J.}\ \bibnamefont
  {Kimble}},\ }\bibfield  {title} {\bibinfo {title} {Dense coding for
  continuous variables},\ }\href {https://doi.org/10.1103/PhysRevA.61.042302}
  {\bibfield  {journal} {\bibinfo  {journal} {Phys. Rev. A}\ }\textbf {\bibinfo
  {volume} {61}},\ \bibinfo {pages} {042302} (\bibinfo {year}
  {2000})}\BibitemShut {NoStop}%
\bibitem [{\citenamefont {Li}\ \emph {et~al.}(2002)\citenamefont {Li},
  \citenamefont {Pan}, \citenamefont {Jing}, \citenamefont {Zhang},
  \citenamefont {Xie},\ and\ \citenamefont {Peng}}]{li2002quantum}%
  \BibitemOpen
  \bibfield  {author} {\bibinfo {author} {\bibfnamefont {X.}~\bibnamefont
  {Li}}, \bibinfo {author} {\bibfnamefont {Q.}~\bibnamefont {Pan}}, \bibinfo
  {author} {\bibfnamefont {J.}~\bibnamefont {Jing}}, \bibinfo {author}
  {\bibfnamefont {J.}~\bibnamefont {Zhang}}, \bibinfo {author} {\bibfnamefont
  {C.}~\bibnamefont {Xie}},\ and\ \bibinfo {author} {\bibfnamefont
  {K.}~\bibnamefont {Peng}},\ }\bibfield  {title} {\bibinfo {title} {Quantum
  dense coding exploiting a bright einstein-podolsky-rosen beam},\ }\href
  {https://doi.org/10.1103/PhysRevLett.88.047904} {\bibfield  {journal}
  {\bibinfo  {journal} {Phys. Rev. Lett.}\ }\textbf {\bibinfo {volume} {88}},\
  \bibinfo {pages} {047904} (\bibinfo {year} {2002})}\BibitemShut {NoStop}%
\bibitem [{\citenamefont {Steinlechner}\ \emph {et~al.}(2013)\citenamefont
  {Steinlechner}, \citenamefont {Bauchrowitz}, \citenamefont {Meinders},
  \citenamefont {M{\"u}ller-Ebhardt}, \citenamefont {Danzmann},\ and\
  \citenamefont {Schnabel}}]{steinlechner2013quantum}%
  \BibitemOpen
  \bibfield  {author} {\bibinfo {author} {\bibfnamefont {S.}~\bibnamefont
  {Steinlechner}}, \bibinfo {author} {\bibfnamefont {J.}~\bibnamefont
  {Bauchrowitz}}, \bibinfo {author} {\bibfnamefont {M.}~\bibnamefont
  {Meinders}}, \bibinfo {author} {\bibfnamefont {H.}~\bibnamefont
  {M{\"u}ller-Ebhardt}}, \bibinfo {author} {\bibfnamefont {K.}~\bibnamefont
  {Danzmann}},\ and\ \bibinfo {author} {\bibfnamefont {R.}~\bibnamefont
  {Schnabel}},\ }\bibfield  {title} {\bibinfo {title} {Quantum-dense
  metrology},\ }\href {https://doi.org/10.1038/nphoton.2013.150} {\bibfield
  {journal} {\bibinfo  {journal} {Nature Photonics}\ }\textbf {\bibinfo
  {volume} {7}},\ \bibinfo {pages} {626} (\bibinfo {year} {2013})}\BibitemShut
  {NoStop}%
\bibitem [{\citenamefont {Liu}\ \emph {et~al.}(2018)\citenamefont {Liu},
  \citenamefont {Li}, \citenamefont {Cui}, \citenamefont {Huo}, \citenamefont
  {Assad}, \citenamefont {Li},\ and\ \citenamefont {Ou}}]{liu2018loss}%
  \BibitemOpen
  \bibfield  {author} {\bibinfo {author} {\bibfnamefont {Y.}~\bibnamefont
  {Liu}}, \bibinfo {author} {\bibfnamefont {J.}~\bibnamefont {Li}}, \bibinfo
  {author} {\bibfnamefont {L.}~\bibnamefont {Cui}}, \bibinfo {author}
  {\bibfnamefont {N.}~\bibnamefont {Huo}}, \bibinfo {author} {\bibfnamefont
  {S.~M.}\ \bibnamefont {Assad}}, \bibinfo {author} {\bibfnamefont
  {X.}~\bibnamefont {Li}},\ and\ \bibinfo {author} {\bibfnamefont
  {Z.}~\bibnamefont {Ou}},\ }\bibfield  {title} {\bibinfo {title}
  {Loss-tolerant quantum dense metrology with su(1,1) interferometer},\ }\href
  {https://doi.org/10.1364/OE.26.027705} {\bibfield  {journal} {\bibinfo
  {journal} {Optics Express}\ }\textbf {\bibinfo {volume} {26}},\ \bibinfo
  {pages} {27705} (\bibinfo {year} {2018})}\BibitemShut {NoStop}%
\bibitem [{\citenamefont {Du}\ \emph {et~al.}(2020)\citenamefont {Du},
  \citenamefont {Chen}, \citenamefont {Ou},\ and\ \citenamefont
  {Zhang}}]{du2020quantum}%
  \BibitemOpen
  \bibfield  {author} {\bibinfo {author} {\bibfnamefont {W.}~\bibnamefont
  {Du}}, \bibinfo {author} {\bibfnamefont {J.}~\bibnamefont {Chen}}, \bibinfo
  {author} {\bibfnamefont {Z.}~\bibnamefont {Ou}},\ and\ \bibinfo {author}
  {\bibfnamefont {W.}~\bibnamefont {Zhang}},\ }\bibfield  {title} {\bibinfo
  {title} {Quantum dense metrology by an su(2)-in-su(1,1) nested
  interferometer},\ }\href {https://doi.org/10.1063/5.0012304} {\bibfield
  {journal} {\bibinfo  {journal} {Applied Physics Letters}\ }\textbf {\bibinfo
  {volume} {117}} (\bibinfo {year} {2020})}\BibitemShut {NoStop}%
\bibitem [{\citenamefont {Li}\ \emph {et~al.}(2023)\citenamefont {Li},
  \citenamefont {Cheng}, \citenamefont {Wang}, \citenamefont {Zhao},
  \citenamefont {Hou}, \citenamefont {Li}, \citenamefont {Rehan}, \citenamefont
  {Zhu}, \citenamefont {Yan}, \citenamefont {Qin} \emph {et~al.}}]{ustc2023}%
  \BibitemOpen
  \bibfield  {author} {\bibinfo {author} {\bibfnamefont {Y.}~\bibnamefont
  {Li}}, \bibinfo {author} {\bibfnamefont {X.}~\bibnamefont {Cheng}}, \bibinfo
  {author} {\bibfnamefont {L.}~\bibnamefont {Wang}}, \bibinfo {author}
  {\bibfnamefont {X.}~\bibnamefont {Zhao}}, \bibinfo {author} {\bibfnamefont
  {W.}~\bibnamefont {Hou}}, \bibinfo {author} {\bibfnamefont {Y.}~\bibnamefont
  {Li}}, \bibinfo {author} {\bibfnamefont {K.}~\bibnamefont {Rehan}}, \bibinfo
  {author} {\bibfnamefont {M.}~\bibnamefont {Zhu}}, \bibinfo {author}
  {\bibfnamefont {L.}~\bibnamefont {Yan}}, \bibinfo {author} {\bibfnamefont
  {X.}~\bibnamefont {Qin}}, \emph {et~al.},\ }\bibfield  {title} {\bibinfo
  {title} {Multi-parameter quantum metrology with stabilized multi-mode
  squeezed state},\ }\href {https://arxiv.org/abs/2312.10379} {\bibfield
  {journal} {\bibinfo  {journal} {arXiv preprint arXiv:2312.10379}\ } (\bibinfo
  {year} {2023})}\BibitemShut {NoStop}%
\bibitem [{\citenamefont {Metzner}\ \emph {et~al.}(2023)\citenamefont
  {Metzner}, \citenamefont {Quinn}, \citenamefont {Brudney}, \citenamefont
  {Moore}, \citenamefont {Burd}, \citenamefont {Wineland},\ and\ \citenamefont
  {Allcock}}]{Oregon2023}%
  \BibitemOpen
  \bibfield  {author} {\bibinfo {author} {\bibfnamefont {J.}~\bibnamefont
  {Metzner}}, \bibinfo {author} {\bibfnamefont {A.}~\bibnamefont {Quinn}},
  \bibinfo {author} {\bibfnamefont {S.}~\bibnamefont {Brudney}}, \bibinfo
  {author} {\bibfnamefont {I.}~\bibnamefont {Moore}}, \bibinfo {author}
  {\bibfnamefont {S.}~\bibnamefont {Burd}}, \bibinfo {author} {\bibfnamefont
  {D.}~\bibnamefont {Wineland}},\ and\ \bibinfo {author} {\bibfnamefont
  {D.}~\bibnamefont {Allcock}},\ }\bibfield  {title} {\bibinfo {title}
  {Two-mode squeezing and su (1, 1) interferometry with trapped ions},\ }\href
  {https://arxiv.org/abs/2312.10847} {\bibfield  {journal} {\bibinfo  {journal}
  {arXiv preprint arXiv:2312.10847}\ } (\bibinfo {year} {2023})}\BibitemShut
  {NoStop}%
%
%
 \bibitem [{Sup()}]{supp}%
   \BibitemOpen
   \bibinfo {note} {See Supplemental Material, which
includes Refs.~\cite{law1998quantum,gardiner2004quantum,blakie2008dynamics,liu2022nonlinear,mao2023quantum,gross2011atomic,kunkel2019simultaneous}, 
for details of the numerical simulation performed using the truncated Wigner method that incorporates imperfections, 
the experimental protocols for simultaneously measuring the observables $Q_{xz}$ and $Q_{yz}$ via a well-designed microwave pulse sequence, 
and the analysis of the metrological gain of our protocol.}\BibitemShut {Stop}%
%
%
\bibitem [{\citenamefont {Gardiner}\ and\ \citenamefont
  {Zoller}(2004)}]{gardiner2004quantum}%
  \BibitemOpen
  \bibfield  {author} {\bibinfo {author} {\bibfnamefont {C.}~\bibnamefont
  {Gardiner}}\ and\ \bibinfo {author} {\bibfnamefont {P.}~\bibnamefont
  {Zoller}},\ }\href@noop {} {\emph {\bibinfo {title} {Quantum noise: a
  handbook of Markovian and non-Markovian quantum stochastic methods with
  applications to quantum optics}}}\ (\bibinfo  {publisher} {Springer Science
  \& Business Media},\ \bibinfo {year} {2004})\BibitemShut {NoStop}%
\bibitem [{\citenamefont {Blakie}\ \emph {et~al.}(2008)\citenamefont {Blakie},
  \citenamefont {Bradley}, \citenamefont {Davis}, \citenamefont {Ballagh},\
  and\ \citenamefont {Gardiner}}]{blakie2008dynamics}%
  \BibitemOpen
  \bibfield  {author} {\bibinfo {author} {\bibfnamefont {P.~B.}\ \bibnamefont
  {Blakie}}, \bibinfo {author} {\bibfnamefont {A.}~\bibnamefont {Bradley}},
  \bibinfo {author} {\bibfnamefont {M.}~\bibnamefont {Davis}}, \bibinfo
  {author} {\bibfnamefont {R.}~\bibnamefont {Ballagh}},\ and\ \bibinfo {author}
  {\bibfnamefont {C.}~\bibnamefont {Gardiner}},\ }\bibfield  {title} {\bibinfo
  {title} {Dynamics and statistical mechanics of ultra-cold bose gases using
  c-field techniques},\ }\href{https://doi.org/10.1080/00018730802564254} {\bibfield  {journal} {\bibinfo
  {journal} {Advances in Physics}\ }\textbf {\bibinfo {volume} {57}},\ \bibinfo
  {pages} {363} (\bibinfo {year} {2008})}\BibitemShut {NoStop}%
%
%
\bibitem [{\citenamefont {Hamley}\ \emph {et~al.}(2012)\citenamefont {Hamley},
  \citenamefont {Gerving}, \citenamefont {Hoang}, \citenamefont {Bookjans},\
  and\ \citenamefont {Chapman}}]{hamley2012spin}%
  \BibitemOpen
  \bibfield  {author} {\bibinfo {author} {\bibfnamefont {C.~D.}\ \bibnamefont
  {Hamley}}, \bibinfo {author} {\bibfnamefont {C.}~\bibnamefont {Gerving}},
  \bibinfo {author} {\bibfnamefont {T.}~\bibnamefont {Hoang}}, \bibinfo
  {author} {\bibfnamefont {E.}~\bibnamefont {Bookjans}},\ and\ \bibinfo
  {author} {\bibfnamefont {M.~S.}\ \bibnamefont {Chapman}},\ }\bibfield
  {title} {\bibinfo {title} {Spin-nematic squeezed vacuum in a quantum gas},\
  }\href {https://doi.org/10.1038/nphys2245} {\bibfield  {journal} {\bibinfo
  {journal} {Nature Physics}\ }\textbf {\bibinfo {volume} {8}},\ \bibinfo
  {pages} {305} (\bibinfo {year} {2012})}\BibitemShut {NoStop}%
\bibitem [{\citenamefont {Gross}\ \emph {et~al.}(2011)\citenamefont {Gross},
  \citenamefont {Strobel}, \citenamefont {Nicklas}, \citenamefont {Zibold},
  \citenamefont {Bar-Gill}, \citenamefont {Kurizki},\ and\ \citenamefont
  {Oberthaler}}]{gross2011atomic}%
  \BibitemOpen
  \bibfield  {author} {\bibinfo {author} {\bibfnamefont {C.}~\bibnamefont
  {Gross}}, \bibinfo {author} {\bibfnamefont {H.}~\bibnamefont {Strobel}},
  \bibinfo {author} {\bibfnamefont {E.}~\bibnamefont {Nicklas}}, \bibinfo
  {author} {\bibfnamefont {T.}~\bibnamefont {Zibold}}, \bibinfo {author}
  {\bibfnamefont {N.}~\bibnamefont {Bar-Gill}}, \bibinfo {author}
  {\bibfnamefont {G.}~\bibnamefont {Kurizki}},\ and\ \bibinfo {author}
  {\bibfnamefont {M.}~\bibnamefont {Oberthaler}},\ }\bibfield  {title}
  {\bibinfo {title} {Atomic homodyne detection of continuous-variable entangled
  twin-atom states},\ }\href {https://doi.org/10.1038/nature10654} {\bibfield
  {journal} {\bibinfo  {journal} {Nature}\ }\textbf {\bibinfo {volume} {480}},\
  \bibinfo {pages} {219} (\bibinfo {year} {2011})}\BibitemShut {NoStop}%
\bibitem [{\citenamefont {Peise}\ \emph {et~al.}(2015)\citenamefont {Peise},
  \citenamefont {Kruse}, \citenamefont {Lange}, \citenamefont {L{\"u}cke},
  \citenamefont {Pezz{\`e}}, \citenamefont {Arlt}, \citenamefont {Ertmer},
  \citenamefont {Hammerer}, \citenamefont {Santos}, \citenamefont {Smerzi}
  \emph {et~al.}}]{peise2015satisfying}%
  \BibitemOpen
  \bibfield  {author} {\bibinfo {author} {\bibfnamefont {J.}~\bibnamefont
  {Peise}}, \bibinfo {author} {\bibfnamefont {I.}~\bibnamefont {Kruse}},
  \bibinfo {author} {\bibfnamefont {K.}~\bibnamefont {Lange}}, \bibinfo
  {author} {\bibfnamefont {B.}~\bibnamefont {L{\"u}cke}}, \bibinfo {author}
  {\bibfnamefont {L.}~\bibnamefont {Pezz{\`e}}}, \bibinfo {author}
  {\bibfnamefont {J.}~\bibnamefont {Arlt}}, \bibinfo {author} {\bibfnamefont
  {W.}~\bibnamefont {Ertmer}}, \bibinfo {author} {\bibfnamefont
  {K.}~\bibnamefont {Hammerer}}, \bibinfo {author} {\bibfnamefont
  {L.}~\bibnamefont {Santos}}, \bibinfo {author} {\bibfnamefont
  {A.}~\bibnamefont {Smerzi}}, \emph {et~al.},\ }\bibfield  {title} {\bibinfo
  {title} {Satisfying the einstein--podolsky--rosen criterion with massive
  particles},\ }\href {https://doi.org/10.1038/ncomms9984} {\bibfield
  {journal} {\bibinfo  {journal} {Nature Communications}\ }\textbf {\bibinfo
  {volume} {6}},\ \bibinfo {pages} {8984} (\bibinfo {year} {2015})}\BibitemShut
  {NoStop}%
\bibitem [{\citenamefont {Liu}\ \emph {et~al.}(2022)\citenamefont {Liu},
  \citenamefont {Wu}, \citenamefont {Cao}, \citenamefont {Mao}, \citenamefont
  {Li}, \citenamefont {Guo}, \citenamefont {Tey},\ and\ \citenamefont
  {You}}]{liu2022nonlinear}%
  \BibitemOpen
  \bibfield  {author} {\bibinfo {author} {\bibfnamefont {Q.}~\bibnamefont
  {Liu}}, \bibinfo {author} {\bibfnamefont {L.-N.}\ \bibnamefont {Wu}},
  \bibinfo {author} {\bibfnamefont {J.-H.}\ \bibnamefont {Cao}}, \bibinfo
  {author} {\bibfnamefont {T.-W.}\ \bibnamefont {Mao}}, \bibinfo {author}
  {\bibfnamefont {X.-W.}\ \bibnamefont {Li}}, \bibinfo {author} {\bibfnamefont
  {S.-F.}\ \bibnamefont {Guo}}, \bibinfo {author} {\bibfnamefont {M.~K.}\
  \bibnamefont {Tey}},\ and\ \bibinfo {author} {\bibfnamefont {L.}~\bibnamefont
  {You}},\ }\bibfield  {title} {\bibinfo {title} {Nonlinear interferometry
  beyond classical limit enabled by cyclic dynamics},\ }\href
  {https://doi.org/10.1038/s41567-021-01441-7} {\bibfield  {journal} {\bibinfo
  {journal} {Nature Physics}\ }\textbf {\bibinfo {volume} {18}},\ \bibinfo
  {pages} {167} (\bibinfo {year} {2022})}\BibitemShut {NoStop}%
%
\bibitem [{\citenamefont {Law}\ \emph {et~al.}(1998)\citenamefont {Law},
  \citenamefont {Pu},\ and\ \citenamefont {Bigelow}}]{law1998quantum}%
  \BibitemOpen
  \bibfield  {author} {\bibinfo {author} {\bibfnamefont {C.~K.}\ \bibnamefont
  {Law}}, \bibinfo {author} {\bibfnamefont {H.}~\bibnamefont {Pu}},\ and\
  \bibinfo {author} {\bibfnamefont {N.~P.}\ \bibnamefont {Bigelow}},\
  }\bibfield  {title} {\bibinfo {title} {Quantum spins mixing in spinor
  bose-einstein condensates},\ }\href
  {https://doi.org/10.1103/PhysRevLett.81.5257} {\bibfield  {journal} {\bibinfo
   {journal} {Phys. Rev. Lett.}\ }\textbf {\bibinfo {volume} {81}},\ \bibinfo
  {pages} {5257} (\bibinfo {year} {1998})}\BibitemShut {NoStop}%
\bibitem [{\citenamefont {Gerbier}\ \emph {et~al.}(2006)\citenamefont
  {Gerbier}, \citenamefont {Widera}, \citenamefont {F\"olling}, \citenamefont
  {Mandel},\ and\ \citenamefont {Bloch}}]{gerbier2006resonant}%
  \BibitemOpen
  \bibfield  {author} {\bibinfo {author} {\bibfnamefont {F.}~\bibnamefont
  {Gerbier}}, \bibinfo {author} {\bibfnamefont {A.}~\bibnamefont {Widera}},
  \bibinfo {author} {\bibfnamefont {S.}~\bibnamefont {F\"olling}}, \bibinfo
  {author} {\bibfnamefont {O.}~\bibnamefont {Mandel}},\ and\ \bibinfo {author}
  {\bibfnamefont {I.}~\bibnamefont {Bloch}},\ }\bibfield  {title} {\bibinfo
  {title} {Resonant control of spin dynamics in ultracold quantum gases by
  microwave dressing},\ }\href {https://doi.org/10.1103/PhysRevA.73.041602}
  {\bibfield  {journal} {\bibinfo  {journal} {Phys. Rev. A}\ }\textbf {\bibinfo
  {volume} {73}},\ \bibinfo {pages} {041602} (\bibinfo {year}
  {2006})}\BibitemShut {NoStop}%
\bibitem [{\citenamefont {Linnemann}\ \emph {et~al.}(2016)\citenamefont
  {Linnemann}, \citenamefont {Strobel}, \citenamefont {Muessel}, \citenamefont
  {Schulz}, \citenamefont {Lewis-Swan}, \citenamefont {Kheruntsyan},\ and\
  \citenamefont {Oberthaler}}]{linnemann2016quantum}%
  \BibitemOpen
  \bibfield  {author} {\bibinfo {author} {\bibfnamefont {D.}~\bibnamefont
  {Linnemann}}, \bibinfo {author} {\bibfnamefont {H.}~\bibnamefont {Strobel}},
  \bibinfo {author} {\bibfnamefont {W.}~\bibnamefont {Muessel}}, \bibinfo
  {author} {\bibfnamefont {J.}~\bibnamefont {Schulz}}, \bibinfo {author}
  {\bibfnamefont {R.~J.}\ \bibnamefont {Lewis-Swan}}, \bibinfo {author}
  {\bibfnamefont {K.~V.}\ \bibnamefont {Kheruntsyan}},\ and\ \bibinfo {author}
  {\bibfnamefont {M.~K.}\ \bibnamefont {Oberthaler}},\ }\bibfield  {title}
  {\bibinfo {title} {Quantum-enhanced sensing based on time reversal of
  nonlinear dynamics},\ }\href {https://doi.org/10.1103/PhysRevLett.117.013001}
  {\bibfield  {journal} {\bibinfo  {journal} {Phys. Rev. Lett.}\ }\textbf
  {\bibinfo {volume} {117}},\ \bibinfo {pages} {013001} (\bibinfo {year}
  {2016})}\BibitemShut {NoStop}%
\bibitem [{\citenamefont {Walls}\ and\ \citenamefont
  {Milburn}(2008)}]{walls2008input}%
  \BibitemOpen
  \bibfield  {author} {\bibinfo {author} {\bibfnamefont {D.}~\bibnamefont
  {Walls}}\ and\ \bibinfo {author} {\bibfnamefont {G.~J.}\ \bibnamefont
  {Milburn}},\ }\bibfield  {title} {\bibinfo {title} {Input--output formulation
  of optical cavities},\ }in\ \href@noop {} {\emph {\bibinfo {booktitle}
  {Quantum optics}}}\ (\bibinfo  {publisher} {Springer},\ \bibinfo {year}
  {2008})\ pp.\ \bibinfo {pages} {127--141}\BibitemShut {NoStop}%
\bibitem [{\citenamefont {Kunkel}\ \emph {et~al.}(2019)\citenamefont {Kunkel},
  \citenamefont {Pr\"ufer}, \citenamefont {Lannig}, \citenamefont
  {Rosa-Medina}, \citenamefont {Bonnin}, \citenamefont {G\"arttner},
  \citenamefont {Strobel},\ and\ \citenamefont
  {Oberthaler}}]{kunkel2019simultaneous}%
  \BibitemOpen
  \bibfield  {author} {\bibinfo {author} {\bibfnamefont {P.}~\bibnamefont
  {Kunkel}}, \bibinfo {author} {\bibfnamefont {M.}~\bibnamefont {Pr\"ufer}},
  \bibinfo {author} {\bibfnamefont {S.}~\bibnamefont {Lannig}}, \bibinfo
  {author} {\bibfnamefont {R.}~\bibnamefont {Rosa-Medina}}, \bibinfo {author}
  {\bibfnamefont {A.}~\bibnamefont {Bonnin}}, \bibinfo {author} {\bibfnamefont
  {M.}~\bibnamefont {G\"arttner}}, \bibinfo {author} {\bibfnamefont
  {H.}~\bibnamefont {Strobel}},\ and\ \bibinfo {author} {\bibfnamefont {M.~K.}\
  \bibnamefont {Oberthaler}},\ }\bibfield  {title} {\bibinfo {title}
  {Simultaneous readout of noncommuting collective spin observables beyond the
  standard quantum limit},\ }\href
  {https://doi.org/10.1103/PhysRevLett.123.063603} {\bibfield  {journal}
  {\bibinfo  {journal} {Phys. Rev. Lett.}\ }\textbf {\bibinfo {volume} {123}},\
  \bibinfo {pages} {063603} (\bibinfo {year} {2019})}\BibitemShut {NoStop}%
\bibitem [{\citenamefont {Mao}\ \emph {et~al.}(2023)\citenamefont {Mao},
  \citenamefont {Liu}, \citenamefont {Li}, \citenamefont {Cao}, \citenamefont
  {Chen}, \citenamefont {Xu}, \citenamefont {Tey}, \citenamefont {Huang},\ and\
  \citenamefont {You}}]{mao2023quantum}%
  \BibitemOpen
  \bibfield  {author} {\bibinfo {author} {\bibfnamefont {T.-W.}\ \bibnamefont
  {Mao}}, \bibinfo {author} {\bibfnamefont {Q.}~\bibnamefont {Liu}}, \bibinfo
  {author} {\bibfnamefont {X.-W.}\ \bibnamefont {Li}}, \bibinfo {author}
  {\bibfnamefont {J.-H.}\ \bibnamefont {Cao}}, \bibinfo {author} {\bibfnamefont
  {F.}~\bibnamefont {Chen}}, \bibinfo {author} {\bibfnamefont {W.-X.}\
  \bibnamefont {Xu}}, \bibinfo {author} {\bibfnamefont {M.~K.}\ \bibnamefont
  {Tey}}, \bibinfo {author} {\bibfnamefont {Y.-X.}\ \bibnamefont {Huang}},\
  and\ \bibinfo {author} {\bibfnamefont {L.}~\bibnamefont {You}},\ }\bibfield
  {title} {\bibinfo {title} {Quantum-enhanced sensing by echoing spin-nematic
  squeezing in atomic bose--einstein condensate},\ }\href
  {https://doi.org/10.1038/s41567-023-02168-3} {\bibfield  {journal} {\bibinfo
  {journal} {Nature Physics}\ }\textbf {\bibinfo {volume} {19}},\ \bibinfo
  {pages} {1585} (\bibinfo {year} {2023})}\BibitemShut {NoStop}%
\bibitem [{\citenamefont {Li}(2021)}]{li2021quantum}%
  \BibitemOpen
  \bibfield  {author} {\bibinfo {author} {\bibfnamefont {X.}~\bibnamefont
  {Li}},\ }\emph {\bibinfo {title} {Quantum enhanced metrology for simultaneous
  multi­parameter estimation}},\ \href@noop {} {Ph.D. thesis},\ \bibinfo
  {school} {Tsinghua University} (\bibinfo {year} {2021})\BibitemShut {NoStop}%
\end{thebibliography}
\end{document}